\documentclass[prb,preprint]{revtex4-1}

\usepackage[latin1]{inputenc}
\usepackage{amsmath,amsfonts,amssymb,mathrsfs}	% additional fonts etc.
\usepackage{graphicx}
\usepackage{verbatim}
\usepackage{color}

\definecolor{ForestGreen}{RGB}{34,139,34}
\definecolor{Cyan}{RGB}{0,191,255}         % this is not the usual cyan, which is very light
\definecolor{Darkgrey}{RGB}{169,169,169}   % *lightest* grey of all three below
\definecolor{Grey}{RGB}{127,127,127}
\definecolor{Dimgrey}{RGB}{105,105,105}    % *darkest* grey of these three ones

\newcommand{\intd}{\textnormal{d}}                  	% ""
\def\intdbar{{\mathchar'26\mkern-12mu \textrm{d}}}	% dbar
\newcommand{\dero}[1]{\frac{\textnormal{d}}{\textnormal{d}#1}}	% d/d...
\newcommand{\diff}[1]{\frac{\textnormal{d}}{\textnormal{d}#1}}	% d/d...
\newcommand{\fracun}[1]{\frac{1}{#1}}  		% fraction "un sur argument"
\newcommand{\undemi}{\frac{1}{2}}  		% 1/2

\newcommand{\skp}[2]{\left\la #1, #2 \right\ra}
\newcommand{\mean}[1]{\left\la #1 \right\ra}    % mean value
\newcommand{\abs}[1]{\left| #1 \right|}
\newcommand{\ie}{i.e.\ }
\newcommand{\eg}{e.g.\ }

\newcommand{\eps}{{\varepsilon}}        %%%%%%%%%%%%%%%%%%%%%%%%%%%%%%%%%%%
\newcommand{\om}{\omega} 
\newcommand{\Om}{\Omega}

\newcommand{\la}{\langle} 
\newcommand{\ra}{\rangle}

\newcommand{\bok}{\mathbf{k}}         %%%%%%%%%%%%%%%%%%%%%%
\newcommand{\bol}{\mathbf{l}}         %%%%%%%%%%%%%%%%%%%%%%
\newcommand{\bop}{\mathbf{p}}         %%%%%%%%%%%%%%%%%%%%%%
\newcommand{\boq}{\mathbf{q}}         %%%%%%%%%%%%%%%%%%%%%%
\newcommand{\bou}{\mathbf{u}}         %%%%%%%%%%%%%%%%%%%%%%
\newcommand{\bzero}{\mathbf{0}}

\newcommand{\cN}{\mathcal{N}}         % Calligraphic Letters %
\newcommand{\cO}{\mathcal{O}}         %                      %
\newcommand{\cT}{\mathcal{T}}

\newcommand{\RR}{\mathbb{R}}            %%%%%%%%%%%%%%%%%%%%%%%%%%%
\newcommand{\NN}{\mathbb{N}}            % Blackboard Bold Letters %
\newcommand{\TT}{\mathbb{T}}
\newcommand{\ZZ}{\mathbb{Z}}
\newcommand{\rIm}{\mathop{\textnormal{Im}}}
\newcommand{\rRe}{\mathop{\textnormal{Re}}}  %%%%%%%%%%%%%%%%%%%%%%%%%%%%%%
\newcommand{\sgn}{\mathop{\mathrm{sgn}}}

\usepackage[english]{babel}
\usepackage{mathtools}                                 %needed for dcases environment

\begin{document}

\title{Self-energy flows in the two-dimensional repulsive Hubbard model}
\author{Kay-Uwe Giering and Manfred Salmhofer\\
Institut f\" ur Theoretische Physik, Universit\" at Heidelberg, Philosophenweg
19, 69120 Heidelberg, Germany}

\date{\today}

\begin{abstract}
We study the two-dimensional repulsive Hubbard model by functional RG methods, 
using our recently proposed channel decomposition of the interaction vertex. 
The main technical advance of this work is that we calculate the full 
Matsubara frequency dependence of the self-energy and the 
interaction vertex in the whole frequency range
without simplifying assumptions on its functional form, and that the effects of 
the self-energy are fully taken into account in the equations for the flow
of the two-body vertex function. At Van Hove filing, we find that the Fermi surface deformations 
remain small at fixed particle density and have a minor impact 
on the structure of the interaction vertex. The frequency dependence 
of the self-energy, however, turns out to be important,
especially 
at a transition from ferromagnetism to $d$-wave superconductivity.
We determine non-Fermi-liquid exponents at this transition point.
\end{abstract}

\maketitle

%\tableofcontents
\newpage

%%%%%%%%%%%%%%%%%%%%%%%%%%%%%%%%%%%%%%%%%%%%%%%%%%%%%%%%%%%%%%%%%%%%%%%%%%%%%%%
\section{Introduction}

The phase diagrams of the cuprates and other currently investigated
materials exhibit a variety of distinct phases, some with symmetry breaking, others
with anomalous normal-state properties. Different effective theories
for the competition of ordering tendencies have been proposed and
compete with each other. On the atomic length scale, the band structure
of these compounds has been determined and a description can be given
by a three-band correlated fermion model.\cite{Emery1987,Markiewicz} 
By further arguments, one can reduce it to a single-band Hubbard model on
a square lattice, 
with a Fermi surface consistent with ARPES measurements.
\cite{Damascelli_ARPES,Kordyuk_ARPES} 
In these models, the electronic Coulomb
interaction is replaced by a short-range repulsion. In this paper we
consider an on-site interaction. This class of itinerant fermion
models was introduced a long time ago, 
\cite{Hubbard63, Kanamori63, Gutzwiller63} but has
received much attention in the context of high-$T_c$ theory, and
is of great theoretical interest by itself, due to the subtleties of
the interplay of the interaction and hopping terms. Major aims are a
controlled derivation of an effective low-energy model for the
cuprates, and in the end, the calculation of the full correlation and
vertex functions of these models. In particular, a full knowledge of
the fermionic self-energy and the two-particle interaction vertex
would allow to predict equilibrium and transport properties, to
compare to Fermi liquid behavior in the normal phase, and to
determine the dynamics of symmetry-broken phases.

In this paper, we continue and conclude our analysis \cite{HusemannSalmhofer2009, HGS2011}
of the functional renormalization group (RG) flow in the level-2-truncation for the 
weakly coupled two-dimensional
Hubbard model in the symmetric phase (all details will be specified below).
In realistic systems the Coulomb interaction is typically not weak, 
and numerous other complications are present. 
However, even the effect of small interactions remains incompletely understood,
and establishing a reliable weak-coupling theory of the Hubbard model 
is therefore an important goal of correlated electron theory. 
Moreover, the approximate phase diagrams obtained at 
weak coupling do resemble those of the true materials, 
so that this may be useful for understanding the cuprates after all. --
We calculate the frequency and momentum dependence of the
fermionic self-energy and the two-body vertex function,
taking the feedback of the self-energy on the vertex flow fully into account. 
For that purpose we solve the functional RG equations
numerically on an adaptive frequency grid, 
hence obtain the frequency dependence without any simplifying assumptions 
on the functional form that it takes. 
We can therefore test the quality of ansatzes made previously, 
which in general turn out
to be accurate only in a very limited frequency range, and we derive improved, 
yet still simple functional forms that are accurate for all but very large frequencies.   
We use this to map out the ordering tendencies in the Hubbard model and 
determine quasiparticle properties in the symmetric phase. 
At Van Hove filling, % (VHF) 
we thereby confirm earlier findings \cite{HonerkampSalmhofer_Tflowletter,HonerkampSalmhofer_Tflow,Hankevych2003} 
of a transition from $d$-wave superconductivity
to ferromagnetism as the ratio of hopping amplitudes between nearest
and next-to-nearest neighbors is varied. 
Our result is again consistent with a quantum critical point at this transition,
and we determine the
exponent for the frequency-dependence of the fermionic self-energy
at this point. This exponent is close, but not
identical, to the exponent found in Ref.\  \onlinecite{Rech2006}; the
deviation may be due to the presence of the Van Hove singularity (VHS)
on the FS in our case.
Because the effective interaction remains small down to very low scales near the
quantum critical point, our results also allow us to give
reliable values for parameters of a detailed effective model in its
vicinity. An analysis of this effective model 
using RG flows with bosons and fermions is in progress.

We now briefly outline the background of the method and the context of our results.
The functional RG gives a functional differential equation for the
correlation functions as a function of an energy or length scale,
%referred to as 
the RG scale. It is a controlled method at weak
coupling\cite{Salmhofer_ContRenFerm}
and has been instrumental in analyzing the competition of
ordering tendencies, 
\cite{ZanchiSchulz98,HalbothMetzner,HonerkampSalmhofer_Umklapp,HonerkampSalmhofer_Tflowletter,HonerkampSalmhofer_Tflow} 
transport properties of systems with impurities 
\cite{KarraschHeddenMeden2008}
as well as 
non-equilibrium phenomena.\cite{SeverinPletyukovSchoeller2010, KarraschEtal2010}
For a recent review, see Ref.~\onlinecite{FRG_Overview}.
In fermionic theories, the functional RG
equation is equivalent to an infinite hierarchy of differential
equations for the vertex functions. The main challenges in solving
this hierarchy are (1) that the vertex functions have a nontrivial
momentum and frequency dependence, which is crucial for understanding
the above-mentioned physical phenomena, and (2) that all ordering
tendencies that turn up
via continuous phase transitions are at some point associated with
interactions that get long-range, hence the momentum- and
frequency-dependent vertex functions get large and eventually singular
at certain loci in momentum space as the RG scale is lowered.
Our present work is a certain completion of addressing the first
challenge in the scale regime where no long-range order has developed,
namely we solve the RG flow equations without any further assumption
on the frequency dependence, and with an already well-tested
approximation for the momentum dependence.
Obtaining an accurate frequency-momentum dependence is particularly
important for quantitative calculations in the 1PI scheme \cite{SalmhoferHonerkamp_ProgTheor} 
where full propagators are always
supported at all scales above the RG scale; in fact, the
$\Omega$-scheme\cite{HusemannSalmhofer2009} employed here
uses only a very mild regulator, for which the propagators are nonzero
everywhere in momentum space and also at all nonzero frequencies.
Schemes with too strong regulators, such as strict momentum-space cutoffs, 
fail to capture ferromagnetic correlations.\cite{HonerkampSalmhofer_Tflow}
The second major challenge mentioned above has been tackled in various papers using
simpler approximations. To get accurate statements about the given
microscopic model, both methods have to be combined, and we now
discuss our approach to this.

From the point of view of RG analysis, one can distinguish different
regimes of energy scales -- the ``high-energy'' regime of degrees of
freedom far away from the Fermi surface (FS); the intermediate
energies, where renormalization effects become sharper and order
parameters start to emerge, i.e.\ selected couplings start to grow in
the flow and self-energy effects become important; 
and the low-energy regime where symmetry breaking happens, gaps may open
on (part of) the FS, and where order parameter averages and
fluctuations play a central role. 
The distinction between these
regimes can always be made conceptually in weakly coupled systems,
but it is particularly pronounced in very weakly coupled systems,
where the non-perturbative symmetry-breaking phenomena generically
occur at scales that are exponentially small in the inverse
interaction strength, but self-energy effects arise already in
low-order perturbation theory and may have important physical effects
at temperatures above symmetry-breaking scales. In special parameter
regimes, the growth of the flowing couplings may be suppressed
altogether and hence the second regime can reach to very low scales;
at quantum critical points, it can reach to scale zero. The fermionic
RG has been used to access the symmetry-broken phase, 
\cite{RGbrokenSym,GerschEtal2005} 
but the full dynamics of the Goldstone modes has been
captured better in a representation with bosonic fields. 
\cite{StrackGerschMetzner08, BaierBickWetterich04} 
On the other hand, the fermionic RG has been more accurate
at the higher and intermediate scales, and the ansatzes made in
bosonic studies have indeed relied strongly on the results of previously done
fermionic flows.

In Ref.~\onlinecite{HusemannSalmhofer2009}, we have proposed a parametrization of the
fermionic RG by a natural channel decomposition, which allows to keep
most details of the frequency-momentum dependence and to switch to a
bosonic description at a certain scale, using the semi-group property
of the RG to stop the flow and a Hubbard-Stratonovich (HS)
transformation to bosonize selected fermionic bilinears, 
namely those corresponding to the most important terms in the 
effective interaction. 
An advantage of this procedure is that the parametrization leaves
little or no ambiguity in the HS transformation of the effective
interaction. Moreover, the parametrization is numerically efficient, 
so that a large number of interaction terms can be kept, to test the accuracy.
To avoid any bias in the flow, the bare on-site
interaction was kept as a separate term in the parametrization of 
Ref.~\onlinecite{HusemannSalmhofer2009}. 
This was then also adopted in a study using boson fields.\cite{Friederich2011} 
The fermionic
formulation furthermore has the advantage of allowing for terms that do not have a
straightforward rewriting in terms of HS fields, hence may be
missed in ansatzes for bosonic actions. One of the results
reported in Ref.~\onlinecite{HGS2011} are interaction terms of exactly this
type, which were discovered only in the fermionic scheme and which
cannot be neglected in the flow. More generally, a restriction to (anti)ferromagnetic 
and superconducting correlations does not describe the vertex function 
of the Hubbard model quantitatively in the filling regime interesting for the cuprates. 

Several RG studies including self-energy effects have been
performed in the past.
In the situation of a moving FS,
one faces the problem that momentum space cutoffs around the free FS 
cannot provide an effective regularization.
By putting a counterterm function in the quadratic part of the action
and solving the emerging inversion problem, 
a map between the non-interacting and interacting FS 
was constructed and extensively studied \cite{FeldmanEtal2000}
for systems with a regular free FS.
Alternatively, the scale decomposition of the propagator 
can be adjusted dynamically according to the moving FS.\cite{DynAdjProp}
With this method, the stationary self-energy was calculated
for several momenta close to the (moving) FS 
and at fixed particle density
in a parameter region with dominant pairing instability.\cite{HonerkampSalmhofer_Umklapp}
Third, the above FS cutoff problem 
can be circumvented by employing an alternative regularization.
With the temperature flow scheme \cite{HonerkampSalmhofer_Tflow}
and in the ferromagnetic parameter region,
the Fermi surface deformation was examined
at zero and non-zero magnetic field
by fitting
the scale derivative of the frequency-independent self-energy for several momenta 
near the FS to an ansatz with two hopping correction terms.\cite{Katanin_SigmaFM} 
The studies in Refs.~\onlinecite{HonerkampSalmhofer_Umklapp,Katanin_SigmaFM}
show that the FS deformation remains small 
and has a minor impact on the structure of the interaction vertex
at the considered points in parameter space;
the FS tends to become more flat in the superconducting parameter region
and to become more curved in the ferromagnetic parameter region.

The self-energy flow equation in the standard 1PI hierarchy 
combined with a stationary parametrization
of the interaction vertex 
leads to the generation of a frequency-independent self-energy.
In previous studies, 
an approximation to the frequency-dependence of the self-energy was obtained 
by calculating the projection of the vertex flow to zero frequency,
but then inserting the integrated vertex flow equation (with the one-loop frequency dependence)
into the self-energy flow equation.
In this way, the self-energy gets a frequency-dependence 
as in two-loop ``sunset diagrams'',
in which the momentum dependence is that of the flowing vertex function.
This approximation was used to calculate the flow of 
quasi-particle scattering rates,\cite{Honerkamp_ScatRate} 
quasi-particle weights,\cite{Zanchi_QuasiparticleWeight, HonerkampSalmhofer_QuasiparticleWeight, Katanin-TwoLoop}
and to continue to real frequencies \cite{RoheMetzner,KataninKampf2004}
(Ref.~\onlinecite{Zanchi_QuasiparticleWeight} uses the Polchinski scheme,
Ref.~\onlinecite{RoheMetzner} the Wick ordered scheme).
Parts of the self-energy can be given back to the flow of the vertex
by assuming that the self-energy correction to the one-particle dispersion
is of the same order of magnitude as the $Z$ factor, 
and then incorporating this $Z$ factor 
in the propagator.\cite{Zanchi_QuasiparticleWeight, HonerkampSalmhofer_QuasiparticleWeight, Katanin-TwoLoop}
All these studies show an anisotropy in the quasi-particle properties
with considerable effects close to the Van Hove points.
In a setting with partial bosonization,
the imaginary part of the frequency-dependent fermionic self-energy was accounted for
through its values at the two lowest Matsubara frequencies and 
at the Van Hove point.\cite{Friederich2011}

In the present work we use the RG scheme for irreducible vertices,
employing the level-2-truncation and Katanin replacement as described in
Ref.~\onlinecite{FRG_Overview}
as well as the vertex parametrization of Ref.~\onlinecite{HusemannSalmhofer2009} 
with its extension to frequency-dependent vertices given in Ref.~\onlinecite{HGS2011}.
In a first calculation, the stationary self-energy is investigated
by determining the flow of corrections to the hopping amplitudes and chemical potential
of the free system.
This allows for resolution of the moving FS and a study of its influence on the RG flow.
We calculate the flow of the stationary self-energy
at fixed particle density, chosen to be interacting VHF,
in a wide range of parameter values.
In a second calculation we study the frequency-dependent self-energy
by discretization in frequency space.
Because we use the channel decomposition of the interaction
vertex we can go beyond the approximation where a frequency-dependent self-energy
is constructed from a stationary vertex,
and we determine the full frequency dependence of the imaginary part of the self-energy.

%%%%%%%%%%%%%%%%%%%%%%%%%%%%%%%%%%%%%%%%%%%%%%%%%%%%%%%%%%%%%%%%%%%%%%%%%%%%%%%
\section{Fermionic RG setup for the Hubbard model}

We consider electrons on the two-dimensional discrete torus $\Gamma = \ZZ_L^2$,
i.e.\ on a square lattice of sidelength $L \in \NN$ with periodic boundary
conditions, %and at temperature $T > 0$ 
and subject to the Hubbard Hamiltonian
 \begin{equation}
  H 
  =
  \sum_{\substack{\bop \in \Gamma^*\\ s}} \eps( \bop )\ c^+_{\bop, s} c_{\bop, s}
  +
  \frac{U}{\abs{\Gamma}} \sum_{\bok, \bop, \boq \in \Gamma^*}
    c^+_{\bop+\bok,+} c^+_{\boq-\bok, -}    
    c_{\boq, -} c_{\bop, +}
.
\end{equation}
Creation and annihilation operators $c^{(+)}_{\bop, s}$ are associated with 
particles of momentum $\bop \in \Gamma^* = ( \frac{2\pi}{L} \ZZ_L )^2$ and 
spin projection $s \in \{\pm\}$. 
The tight-binding dispersion 
\begin{equation}
  \eps( x, y ) = 
  -2 t_1\ ( \cos x + \cos y ) + 4 t_2\ ( \cos x \cos y + 1 ) - \mu
\end{equation}
describes particles hopping between nearest and next-to-nearest lattice
neighbors. 
We consider ratios $0 < t_2/t_1 < \undemi$. 
Here a chemical potential $\mu$ is already included, and $\mu = 0$ 
corresponds to free VHF.
The screened Coulomb interaction is mimicked by an on-site repulsion $U > 0$.
The model exhibits a non-trivial interplay
of the kinetic part, which is diagonal in momentum space, and the potential part,
which is diagonal in position space.

In the path integral formulation, the action
\begin{equation}\label{eq:Hubbard_action}
  S_\Om( \bar\psi,\psi )
=
  \int\intd p\sum_s q_\Om(p)\ \bar\psi_{ps}\psi_{ps}
 -U \int\intd k\ \intd p\ \intd q\  
     \bar\psi_{p+k, +} \bar\psi_{q-k, -} \psi_{q, -} \psi_{p, +}
\ .
\end{equation}
is a function of Grassmann variables $\bar\psi_{ps}$, $\psi_{ps}$
that are labeled by 
frequency-momentum tuples $p = (p_0, \bop) \in M_n \times \Gamma^*$
and spin projection $s$. 
Fermionic Matsubara frequencies read
$\hat{p}_0 = nT\ ( 1 - e^{-i\pi p_0/n})/i$ with $p_0 \in M_n = \{1, 3, \dotsc, 2n-1\}$
and result from the
division of the interval $[0, 1/T]$ in $n \in \NN$ time slices.
We adopt shorthand notation
$\int\intd p = \frac{T}{\abs{\Gamma}} \sum_{ p_0 \in M_n, \bop \in \Gamma^*}$
and 
$\delta( p-q ) = \frac{\abs{\Gamma}}{T} \delta_{p_0, q_0}\delta_{\bop, \boq}$.
Subsequently, the zero temperature limit $T \to 0$ will be taken.

To do the RG, we use the $\Omega$-scheme, where 
the inverse free propagator is multiplied by a soft frequency
regulator,\cite{HusemannSalmhofer2009}
\begin{equation}\label{eq:Om_regulator}
  q_\Om(p) = \Big( i\hat{p}_0  - \eps_\bop \Big)\ /\ { \chi_\Om( \hat{p}_0 ) },
  \quad\quad\quad 
  \chi_\Om(\om)  = \frac{\om^2}{\om^2 + \Om^2}
.
\end{equation}
The scale parameter $\Om \ge 0$ 
sets a scale in that for $|\omega| \gg \Omega$, the regularized
propagator is very close to the original one, while the pole
of $q_0 (p)^{-1}$ at $p_0 =0$ and $\bop$ on the FS 
gets replaced by a value of $q_\Omega (p)^{-1}$ that is of order $\Omega$.
In the limit $\Om \to \infty$ the free propagator vanishes whereas
for $\Om = 0$ the regularization is removed.
This regulator, as opposed to a momentum shell cutoff,
allows for a simple regularization with moving Fermi surface 
due to self-energy effects. 
Moreover, small-momentum particle-hole processes,
which become important for $t_2 \gtrsim 0.3 t_1$ and VHF, 
are not artificially suppressed.\cite{HonerkampSalmhofer_Tflow}

We use the regularization parameter $\Om$ as the RG scale parameter, and
calculate the fermionic two- and four-point functions within the RG scheme 
for one-particle irreducible (1PI) Green functions,
employing the level-two-truncation.\cite{SalmhoferHonerkamp_ProgTheor,FRG_Overview}
We choose an ansatz for the 1PI generating functional that 
preserves all symmetries (translational, spin SU$(2)$, charge U$(1)$) 
of the action and restrict
subsequent calculations to the symmetric regime. 

The general form of this symmetric 1PI functional is \cite{SalmhoferHonerkamp_ProgTheor}
\begin{multline}\label{eq:1pi-parametrisation}
  \Gamma_\Om( \bar\psi,\psi )
=
  \int \intd p \	\sum_s \ 
\big(
q_\Om (p) + \Sigma_\Om (p)
\big)  \;
\bar\psi_{p, s} \psi_{p, s} \ 
\\
  -\fracun{2^2}
  \int \intd p_1 \dotsc \intd p_4 \ 	\sum_{s_1 \dotsc s_4}
  \delta( p_1 + p_2 - p_3 - p_4 )\ 
  \bar\psi_{p_1, s_1} \bar\psi_{p_2, s_2} \psi_{p_3, s_3} \psi_{p_4, s_4} \ 
\\
  \fracun{2} 
  \Bigl(
  v_\Om( p_1, p_2, p_3, p_4 ) \delta_{s_1s_4} \delta_{s_2s_3}
  -v_\Om( p_1, p_2, p_4, p_3 ) \delta_{s_1s_3} \delta_{s_2s_4}
\Bigr)
.
\end{multline}
Here $\Sigma_\Om$ is the (thermal) self-energy, and $v_\Om$
the two-body interaction vertex. 
We write $v_\Om( p_1, p_2, p_3, p_4 )$ as a function of four
frequency-momentum arguments, 
but by translational symmetry it only depends on three of them. 
Subsequent equations should therefore be read in the subspace $p_1 + p_2 - p_3 - p_4 = 0$.

The RG equations for the above coefficient functions read\cite{SalmhoferHonerkamp_ProgTheor}
\begin{align}\label{eq:Sigma_v_flow}
\dot \Sigma_\Om( p )
&=
\fracun{2}
\int \intd l\
s_\Om(l)\ 
\Bigl(
v_\Om( p, l, p, l) - 2 v_\Om(p, l, l, p)
\Bigr)
,
\nonumber\\
\dot v_\Om ( p_1, \dotsc, p_4 ) 
&=
(\cT_{pp} + \cT_{ph, cr} + \cT_{ph, d})( p_1, \dotsc, p_4 )
,
\end{align}
where
\begin{align}
\cT_{pp} ( p_1, \dotsc, p_4 )
&=
-\fracun{2} \int \intd l\ L_\Om( l, p_1+p_2-l )\ 
v_\Om(p_1, p_2, l, p_1+p_2-l)\  v_\Om(p_1+p_2-l, l, p_3, p_4),
\nonumber\\
\cT_{ph, cr}( p_1, \dotsc, p_4 ) 
&=
-\fracun{2} \int \intd l\ L_\Om( l, p_1-p_3+l  )\ 
v_\Om(p_1, l, p_3, p_1-p_3+l)\  v_\Om(p_1-p_3+l, p_2, l, p_4),
\nonumber\\
\cT_{ph, d}( p_1, \dotsc, p_4 ) 
&=
+\fracun{2} \int \intd l\ L_\Om( l, p_2-p_3+l )\ 
\Bigl(
2 v_\Om(p_1, p_2-p_3+l, l, p_4)\  v_\Om(l, p_2, p_3, p_2-p_3+l)
\nonumber\\
&\qquad\qquad
- v_\Om(p_1, p_2-p_3+l, l, p_4)\  v_\Om(l, p_2, p_2-p_3+l, p_3)
\nonumber\\
&\qquad\qquad
- v_\Om(p_1, p_2-p_3+l, p_4, l)\  v_\Om(l, p_2, p_3, p_2-p_3+l)
\Bigr)
\end{align}
denote the particle-particle, crossed and direct particle-hole contribution, respectively.
The so-called single-scale propagator is given by
$ s_\Om(p) = -g_\Om^2(p) \dot q_\Om(p)$ 
and the full propagator is 
\begin{equation}
g_\Om (p) 
=
\big(
q_\Om (p) + \Sigma_\Om (p)
\big)^{-1}
.
\end{equation}
We replace\cite{Katanin_WardId}
$L_\Om(p_1, p_2) = s_\Om(p_1)g_\Om(p_2) + g_\Om(p_1)s_\Om(p_2) \to \diff{\Om}\ \big( g_\Om(p_1) g_\Om(p_2) \big)$.
By this modification of the 1PI flow equations
the feed-back of parts of the irreducible six-point vertex to $v_\Om$
is included.
This reorganization of flow equations
was found to be crucial for recovering the exact solution
to the reduced BCS model within the RG method.\cite{RGbrokenSym}

In the limit $\Om \to \infty$, the microscopic action \eqref{eq:Hubbard_action} should be recovered. 
This is why we supplement the system of differential equations with initial conditions via imposing
$v_\Om(p_1,\dotsc,p_4) \to 2U$ and
$\Sigma_\Om(p) \to -U/2$ as $\Om\to\infty$.
The non-zero limit of the self-energy for vanishing propagator is a special property of the
$\Om$ regularization, 
$\int\intd l\ q_\Om^{-1}(l) \to_{\Om\to\infty} 1/2$.

Once the flow equations have been derived it is convenient to take the 
limit of continuous time\cite{Salmhofer_Book} $n \to \infty$. 
We work directly in the infinite-volume limit $L \to \infty$, 
and also restrict to zero temperature, hence take $T \to 0$
at $\Om > 0$. 
Frequency-momentum tuples then take values $p = (p_0, \bop) \in \RR \times [-\pi, \pi]^2$,
the inverse free propagator becomes
$ q_\Om(p) = \left( ip_0 - \eps_\bop \right) / { \chi_\Om(p_0) }$, 
and we have $\delta( p-q ) = (2\pi)\delta(p_0 - q_0)\ (2\pi)^2\delta(\bop-\boq)$
and $\int\intd p  = \int\intdbar p_0 \int\intdbar \bop$, 
with
$\int \intdbar p_0 = (2\pi)^{-1} \int_\RR\intd p_0$
and
$\int \intdbar \bop = (2\pi)^{-2} \int_{[-\pi, \pi]^2}\intd \bop$.

At zero temperature and VHF, a flow to strong coupling is observed:
The interaction vertex develops a strong frequency-momentum dependence
with divergences for several frequency-momentum pairs at a non-vanishing scale $\Om > 0$. 
We stop integration of the flow when the interaction vertex exceeds a fixed multiple
of the free bandwidth, 
\ie at $\max\abs{v_\Om} = 40 t_1$. 
This defines a ``stopping scale'' $\Om_*$, 
which has the interpretation of an energy scale where correlations of
particle-particle or particle-hole pairs become important\footnote{The present text and Ref.~\citenum{HGS2011}
differ by a conventional factor of two in the interaction vertex;
the stopping condition is the same in both studies.}.
To continue the flow to lower scales, a non-symmetric parametrization of the 1PI functional
is needed.
The tendency towards a specific ordering is, however, 
already visible from the growth of 
the symmetric interaction vertex.

%%%%%%%%%%%%%%%%%%%%%%%%%%%%%%%%%%%%%%%%%%%%%%%%%%%%%%%%%%%%%%%%%%%%%%%%%%%%%%%%%%%%%%%%%%%%%%%%%%%%%%%%%%%%%%%%
\section{Exchange parametrization for the interaction vertex}

The parametrization for the interaction vertex proposed in 
Ref.~\onlinecite{HusemannSalmhofer2009} is designed to capture 
the most singular vertex structure in a systematic way.  
Although used there only for frequency-independent
vertices, it allows to include both the singular momentum 
\emph{and} frequency dependence of the interaction vertex. 
Here we account also for frequency dependences
and use this parametrization 
to study the RG flow including self-energy effects.

The exchange parametrization is set up as follows:
The solution to the flow equation \eqref{eq:Sigma_v_flow} for the interaction vertex 
can 
be written
$v_\Om(p_1\dotsc p_4) = 2U + 
\left( v_{pp}^\Om + v_{ph,cr}^\Om + v_{ph, d}^\Om \right)(p_1\dotsc p_4)$,
with $\dot v_{pp}^\Om = \cT_{pp}$, $\dot v_{ph, cr}^\Om = \cT_{ph, cr}$,
$\dot v_{ph, d}^\Om = \cT_{ph, d}$.

The differential equation for $v_{pp}^\Om$ has the following structure (equations for $v_{ph,cr}^\Om$ and $v_{ph,d}^\Om$
are treated in the same way): 
The integrand on the right-and side contains the product of two propagators, 
which exhibit singularities for certain values of the loop frequency-momentum, 
\ie $l$ or $p_1+p_2-l$ have zero frequency and zero energy.
Depending on the external frequency-momenta $p_1, \dotsc, p_4$, 
these two poles can coincide.
For such a configuration the flow is strongly driven, 
and it can be expected that this generates the most singular vertex structure.

The feed-in of the interaction vertex itself to the flow is essential because it 
multiplies this propagator product, in particular its values 
near the propagator poles are of importance.
If one assumes that the feed-in
of the interaction vertex does not dominate the above mechanism, 
a simple form for the most singular structure of the interaction vertex can be written down.

Since the external variables $p_1, \dotsc, p_4$ enter the propagator product only via the
linear combination $p_1 + p_2$ the most singular dependence of $v_{pp}^\Om$ on its arguments 
can be expected through the transfer frequency-momentum $p_1+p_2$.
The remaining dependence on external variables should not be neglected since it can
produce an important modulation of this singular structure. 
However, a specific form for this remaining frequency-momentum dependence can be assumed, 
for which previous studies using a Fermi surface 
patching\cite{ZanchiSchulz98,HonerkampSalmhofer_Umklapp,HalbothMetzner}
or full Brillouin zone discretization\cite{HonerkampInteractionFlow2004}
can serve as a guideline,
and which can be tested subsequently.

Interpretation of $v_{pp}^\Om$, $v_{ph,cr}^\Om$, $v_{ph,d}^\Om$ can be found by regarding
the spin structure of the interaction vertex. It turns out that these functions are directly
connected to different interaction channels of two fermions, namely
to interacting Cooper pairs $v_{SC}^\Om$ := $v_{pp}^\Om$,
spin interaction $v_{M}^\Om$ := $v_{ph, cr}^\Om$
and
charge interaction $v_{K}^\Om$ := $2 v_{ph, d}^\Om - T_{34}v_{ph, cr}^\Om$.
$T_{34}$ is the exchange operator defined as
$T_{34} v(p_1,p_2,p_3,p_4) = v(p_1, p_2, p_4, p_3)$.

We continue by using the functions $v_{M}^\Om$, $v_{SC}^\Om$, $v_{K}^\Om$,
then
\begin{equation}\label{eq:v_decomp}
  v_\Om( p_1, \dotsc, p_4)
  =
  2U +
  ( v_{SC}^\Om + v_M^\Om + \fracun{2} v_K^\Om ) ( p_1, \dotsc, p_4)
  + \fracun{2} v_M^\Om( p_1, p_2, p_4, p_3)
.
\end{equation}

Following the above reasoning, each channel is Fourier decomposed
in the non-transfer momenta in a way such that basic vertex symmetries
(see equations \eqref{eq:v_symmetries} below)
are satisfied.
The pairing channel \eg is written
\begin{align}
  v_{SC}^\Om(p_1, \dotsc, p_4)
= &
  D^\Om(p_1 + p_2, \frac{ p_1 + p_2 }{2} - p_1, \frac{p_1 + p_2}{2} - p_3)
\nonumber\\
= &
  \sum_{m, n} f_m(\frac{ \bop_1 +\bop_2}{2} -\bop_1 ) f_n(\frac{ \bop_1 +\bop_2}{2} -\bop_3 )
  D_{mn}^\Om( p_1 + p_2, \frac{ p^0_1 + p^0_2}{2} - p^0_1, \frac{ p^0_1 + p^0_2}{2} - p^0_3)
.
\end{align} 

The full frequency structure of the interaction vertex including frequency-dependent
boson-fermion vertex functions was studied in Ref.~\onlinecite{HGS2011}.
Non-transfer frequency dependences were found to be of minor importance to
the parametrization of the most singular structure of the symmetric
interaction vertex.
For the examination of self-energy effects we thus disregard
non-transfer frequencies here.

This results in rewriting the three interaction channels
\begin{align}\label{eq:exchangeparametrisation}
v_{SC}^\Om ( p_1 \dotsc p_4 )
&=
-
\sum_{m, n} f_m( \frac{ \bop_1 + \bop_2 }{2} - \bop_1 )\ D_{mn}^\Om( p_1 + p_2 )\ f_n( \frac{ \bop_1 + \bop_2 }{2} - \bop_3 ),
\nonumber\\
v_{M}^\Om ( p_1 \dotsc p_4 )
&=
+
\sum_{m, n} f_m( \bop_1 -  \frac{ \bop_1 - \bop_3 }{2} )\ M_{mn}^\Om( p_1 - p_3 )\ f_n( \bop_2 +  \frac{ \bop_1 - \bop_3 }{2} ),
\nonumber\\
v_{K}^\Om ( p_1 \dotsc p_4 )
&=
-
\sum_{m, n} f_m( \bop_1 +  \frac{ \bop_2 - \bop_3 }{2} )\ K_{mn}^\Om( p_2 - p_3 )\ f_n( \bop_2 -  \frac{ \bop_2 - \bop_3 }{2} )
\end{align}
in terms of exchange propagators $\{ M_{mn}^\Om, D_{mn}^\Om, K_{mn}^\Om \}$ and form factors $\{ f_m \}$.
In the following, scale parameter indices will be dropped.

The parametrization \eqref{eq:exchangeparametrisation} does not uniquely determine exchange propagators
since it disregards non-transfer frequencies. 
Equations \eqref{eq:exchangeparametrisation} can be solved for exchange propagators after specification 
of a choice for non-transfer frequencies. We consider these two frequencies as a function of 
the transfer frequency, thus projecting the frequency space to a line.
All frequency projections that map a bosonic frequency to a fermionic one and
respect the vertex symmetries are admitted. 

At temperature $T = 0$, non-transfer frequencies can \eg be chosen as 
half of the transfer frequency,
\begin{align}
  D_{mn}(p)
  &=
  -\int \intdbar \bok_1 \intdbar \bok_3\ 
  f_m( \frac{\bop}{2} - \bok_1)\ f_n( \frac{\bop}{2} - \bok_3)\ 
  v_{SC}( k_1, p-k_1, k_3, p-k_3 ) 
  \Big|_{ k_1^0 = k_3^0 = \frac{p_0}{2} }\ ,
  \nonumber\\
  M_{mn}(p)
  &=
  +\int \intdbar \bok_1 \intdbar \bok_2\ 
  f_m( \bok_1 - \frac{\bop}{2} )\ f_n( \bok_2 + \frac{\bop}{2} )\ 
  v_{M}( k_1, k_2, k_1-p, k_2+p ) 
  \Big|_{ k_1^0 = \frac{p_0}{2}, k_2^0 = -\frac{p_0}{2} }\ ,
  \nonumber\\
  K_{mn}(p)
  &=
  -\int \intdbar \bok_1 \intdbar \bok_2\ 
  f_m( \bok_1 + \frac{\bop}{2} )\ f_n( \bok_2 - \frac{\bop}{2} )\ 
  v_{K}( k_1, k_2, k_2-p, k_1+p ) 
  \Big|_{ k_1^0 = -\frac{p_0}{2}, k_2^0 = \frac{p_0}{2} }
.
\end{align}

In the frequency-independent setup for systems at VHF,
major contributions in the flow with exchange propagators come from
diagonal propagator elements, namely exchange propagators
associated with an $s$-wave form factor 
$f_1(x, y) = 1$
in all channels as well as a $d$-wave form factor
$f_2(x, y) = \cos x - \cos y$ in the superconducting channel.\cite{HusemannSalmhofer2009}
A subsequent study\cite{OrtloffHusemann} shows, by comparison to a flow
with discretization of the interaction vertex itself as a function of three
fermion momenta, that this decomposition captures well the interaction vertex structure 
for several Fermi surface geometries. It also indicates that, in the parameter region
of strong competition between FM and $d$-SC ordering tendencies, the $d$-wave form factor
could be slightly modified, possibly in a scale-dependent way, in order to resolve the full singular
structure of the interaction vertex. This modification of $f_2$ is currently investigated.

\begin{figure}
  \centering
 %  \scalebox{0.65}{
 \includegraphics{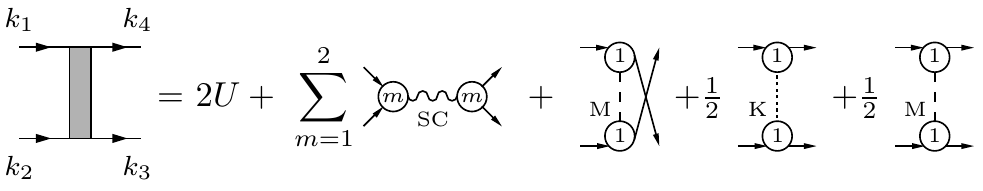}
  
\caption{Graphical representation of the parametrization 
\eqref{eq:v_decomp} and \eqref{eq:exchangeparametrisation}
of the interaction vertex with 
an $s$-wave form factor $f_1(x, y) = 1$ in the three interaction channels
and a $d$-wave form factor $f_2(x, y) = \cos x - \cos y$ in the pairing channel.}
\label{fig:v_decomp}
\end{figure}

We restrict the vertex flow by keeping only the functions
$M_{11}, K_{11}, D_{11}, D_{22}$. The RG equations then read

\begin{align}\label{eq:Sigmadot}
   \dot \Sigma ( p )
   =&
   \fracun{2} ( -2U + K_{11}(0) )\ \int \intd l\ s(l)\nonumber\\
   &
   + \fracun{2} \int \intd l\ s(l-p)\
    \left[ D_{11}(l) + D_{22}(l)\ f_2^2(\frac{\bol}{2}-\bop) \right]
   - \fracun{2^2} \int \intd l\ s(l+p)\
    \left[ K_{11}(l) + 3M_{11}(l) \right]
\end{align}
and
\begin{align}\label{eq:Bmmdot}
   \dot D_{mm}( p )
   &=
   + \fracun{2} \int \intd l\  L\left( -( l - \frac{p}{2} ), l + \frac{p}{2} \right) \ 
   F_m^2\left( -D_{mm}(p), \frac{3M-K}{2} \right) (l_0, \bol, \bop ),
   \quad\quad m = 1,2,
   \nonumber\\
   \dot M_{11}( p )
   &=
   -\fracun{2} \int \intd l\  L\left(  l - \frac{p}{2} , l + \frac{p}{2} \right) \ 
   F_1^2\left( +M_{11}(p), \frac{-2D+M-K}{2} \right) (l_0, \bol, \bop ),
   \nonumber\\
   \dot K_{11}( p )
   &=
   -\fracun{2} \int \intd l\  L\left(  l - \frac{p}{2} , l + \frac{p}{2} \right) \ 
   F_1^2\left( -K_{11}(p), \frac{-2D+3M+K}{2} \right) (l_0, \bol, \bop )
.
\end{align}

The feed-back of the interaction vertex to the flow is given by
\begin{align}\label{eq:define-F_m}
  F_1(A, B)(l_0, \bol, \bop)
  =&
  2U + A
  \nonumber\\
  &
  + \int \intdbar\bou\ 
    \Bigl(
    B_{11}(l_0, \bou) + 
    B_{22}(l_0, \bou)\ f_2( - \frac{\bou}{2} + \bol - \frac{\bop}{2})
    			f_2( - \frac{\bou}{2} + \bol + \frac{\bop}{2})
    \Bigr),
  \nonumber\\
  F_2(A, B)(l_0, \bol, \bop)
  =&
  A\ f_2( \bol  )
  \nonumber\\
  &
  +  \int \intdbar\bou\ B_{11}(l_0, \bou)\ f_2( \bou - \bol  )
,
\end{align}
and $M_{22} \equiv K_{22} \equiv 0$ is understood.

The dependence of functions $F_i$ on their momentum arguments can be written explicitly
 by the help of trigonometric identities,
\begin{align}\label{eq:define-explicit-F_m}
  F_1(A, B)(l_0, \bol, \bop)
  &=
   2U + A
 \nonumber\\
 &+ \mean{ B_{11} }_1(l_0)
 \nonumber\\
& + \mean{ B_{22} }_2(l_0) 
   \left(    \cos( l_x - \frac{p_x}{2} ) \cos( l_x + \frac{p_x}{2} )
          +  \cos( l_y - \frac{p_y}{2} ) \cos( l_y + \frac{p_y}{2} ) 
   \right)
 \nonumber\\
& + \mean{ B_{22} }_3(l_0) 
   \left(    \sin( l_x - \frac{p_x}{2} ) \sin( l_x + \frac{p_x}{2} )
          +  \sin( l_y - \frac{p_y}{2} ) \sin( l_y + \frac{p_y}{2} ) 
   \right)
 \nonumber\\
& - \mean{ B_{22} }_4(l_0) 
   \left(    \cos( l_x - \frac{p_x}{2} ) \cos( l_y + \frac{p_y}{2} )
          +  \cos( l_x + \frac{p_x}{2} ) \cos( l_y - \frac{p_y}{2} ) 
   \right), 
\nonumber\\
 F_2(A, B)(l_0, \bol, \bop)
 =&
 \Big( A  +  \mean{ B_{11} }_5(l_0) \Big)\ 
     f_2( \bol ),
\end{align}
where
$\mean{ B }_i(l_0) =  \int \intdbar \bou\  B( l_0, \bou )\ g_i( \bou )$
and
\begin{align}
  g_1(x, y) &= 1,
  \nonumber\\
  g_2(x, y) &= \undemi \left( \cos^2 \frac{x}{2}  + \cos^2 \frac{y}{2} \right),
  \nonumber\\
  g_3(x, y) &= \undemi \left( \sin^2 \frac{x}{2}  + \sin^2 \frac{y}{2} \right), 
  \nonumber\\  
  g_4(x, y) &=                \cos \frac{x}{2} \cos \frac{y}{2},  
  \nonumber\\  
  g_5(x, y) &= \undemi \left( \cos x  + \cos y \right) 
.
\end{align}

\medskip
We remark that by the symmetries
(i) $F_i(A, B)(l_0, -\bol, \bop) =  F_i(A, B)( l_0, \bol, \bop )$
and
(ii) $L\Big( \pm (l - \frac{p}{2}), l + \frac{p}{2} \Big)
= L\Big( \Big(\pm (l_0 - \frac{p_0}{2} ), \bol - \frac{\bop}{2} \Big), l + \frac{p}{2} \Big)$
of the integrand in \eqref{eq:Bmmdot},
loop momentum integration can be restricted to half of the Brillouin zone (BZ).

\medskip
The essential feed-in of the $d$-wave pairing channel to $s$-wave channels
can be reduced to a simple form,
which produces an approximation to the function $F_1$:
The exchange propagator $D_{22}(p_0, \bop)$, as a function of momentum,
generically exhibits a pronounced maximum at $\bop = \bzero$ and then decays rapidly. 
Thus, the momentum integrals $\mean{D_{22}}_3$ and $\mean{D_{22}}_4 - \mean{D_{22}}_2$
will be small compared to $\mean{D_{22}}_2$. 
Moreover, the
$\mean{D_{22}}_3$ term in $F_1$ is multiplied  with a function that vanishes
at $\bol \pm \fracun{2}\bop = (0, \pi)$, i.e. is suppressed 
near the Van Hove points, which dominate the flow at VHF.
Dropping the $\mean{D_{22}}_3$ and $\mean{D_{22}}_4 - \mean{D_{22}}_2$ terms in $F_1$ yields
the approximation
\begin{equation}\label{eq:F1_approx}
  F_1(A, B)(l_0, \bol, \bop)
  \approx
   2U + A\ 
   + \mean{ B_{11} }_1(l_0)\  
   + \mean{ B_{22} }_2(l_0)\ f_2( \bol - \frac{\bop}{2} ) f_2( \bol + \frac{\bop}{2} )
.
\end{equation}
We have verified in numerous situations that this approximation produces results of high accuracy.
In some setups this approximation can substantially lighten numerical efforts,
it will be applied for the examination of the frequency-dependent self-energy, sec. \ref{sec:ImSigma}.

%%%%%%%%%%%%%%%%%%%%%%%%%%%%%%%%%%%%%%%%%%%%%%%%%%%%%%%%%%%%%%%%%%%%%%%%%%%%%%%%%%%%%%%%%%%%%%%%%%%%%%%%%%%%%%%%%%%%%%%
\section{Symmetry considerations and discretization procedure}

The flow equations constitute a system of differential equations for functions. 
Parametrization of frequency-momentum dependence for vertex functions will reduce the system
to a number of coupled ODEs that can be studied numerically.
This parametrization is guided by symmetry considerations.

\medskip
Let
$S (p_0, \bop) = (-p_0, \bop) $
denote the operator of time reflection
and 
$ T (p_0, \bop) = (p_0, \hat T \bop)$
an operator of spatial transformation. 
Let be 
$\TT = \left\{
{
%\footnotesize
\scriptsize
%\tiny
\left( \begin{array}{c c} -1 & 0 \\ 0 & 1 \end{array} \right),
\left( \begin{array}{c c} 1 & 0 \\ 0 & -1 \end{array} \right),
\left( \begin{array}{c c} 0 & 1 \\ 1 & 0 \end{array} \right)
}
\right\}$
a set of spatial reflections.
The solution to the 
% symmetric 
flow equations \eqref{eq:Sigma_v_flow} 
with the initial condition stated above exhibits the following symmetries
\begin{align}\label{eq:Sgm_symmetries}
 \Sigma( Sp )  &=  \overline{ \Sigma(p) },
 &
 \textrm{ (time reversal symmetry) }
 \nonumber\\
 \Sigma( Tp ) &= \Sigma( p )\ \forall\ \hat T \in \TT
 &
 \textrm{ (spatial reflection symmetries) }
\end{align}
and
\begin{align}\label{eq:v_symmetries}
 v( p_2, p_1, p_4, p_3)  &=  v( p_1, p_2, p_3, p_4),
 &
 \textrm{ (anti-symmetrization property) }
 \nonumber\\
 v( p_3, p_4, p_1, p_2) &=  v( p_1, p_2, p_3, p_4), 
 & 
 \textrm{ (particle-hole symmetry) }
 \nonumber\\
 v( Sp_1, \dotsc, Sp_4 ) &= \overline{ v( p_1, \dotsc, p_4 ) },
 &
 \textrm{ (time reversal symmetry) }
 \nonumber\\
 v( Tp_1, \dotsc, Tp_4 ) &=  v( p_1, \dotsc, p_4 )\ \forall\ \hat T \in \TT
 &
 \textrm{ (spatial reflection symmetries) }.
\end{align}
All these symmetries are a consequence of the 
specific ansatz \eqref{eq:1pi-parametrisation} for the effective action. 
The anti-symmetrization property directly results from the expansion
\eqref{eq:1pi-parametrisation}.
The other symmetries are valid for the initial condition and in addition
compatible with the flow equation, hence hold during the flow.
The spatial reflection symmetries follow from the symmetries of the free
dispersion relation $\eps(x, y)$.

As a consequence, exchange propagators exhibit symmetries
\begin{align}\label{eq:B_symmetries}
  B_{mm}( Sp ) &= \overline{ B_{mm}( p ) },
  &
  \textrm{ (time reversal symmetry) }
  \nonumber\\
  B_{mm}( Tp ) &= B_{mm}( p )\ \ \forall\ \hat T \in \TT
  &
  \textrm{ (spatial reflection symmetries) }
\end{align}
for each of the propagator variables $B = D,\ M,\ K$.
Furthermore, the combination of anti-symmetrization property and particle-hole symmetry implies
$  D^t( p ) = D(  p )$,
$  M^t( p ) = M( -p )$,
$  K^t( p ) = K( -p )$.
Thus, diagonal elements in the magnetic and scattering channel are real,
$M_{mm}(p), K_{mm}(p) \in \RR$.

\begin{figure}
  \centering
  \scalebox{0.65}{
  \includegraphics{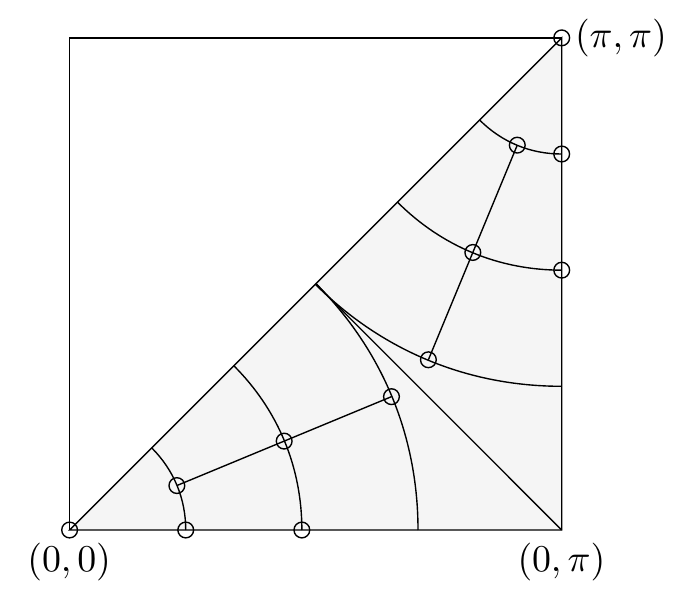}
  }
  
\caption{Illustration of momentum discretization for exchange propagators in 12 segments
about $\bop = (0, 0)$ and $\bop = (\pi, \pi)$. Small circles mark the representative momentum
associated with each segment. 
For detection of incommensurate AFM ordering tendencies 
placement of data points along the coordinate axes is advisable.
}
\label{fig:sector-discretisation}
\end{figure}

\medskip
We continue with the description of our discretization procedure.
The momentum dependence of exchange propagators is approximated by step functions,
as in Ref.\ ~\onlinecite{HusemannSalmhofer2009}.
This means we decompose the Brillouin zone into sectors centered about momenta 
$\bop = (0, 0)$ or $\bop = (\pi, \pi)$
and
designed such that momentum regions with strong variation of exchange propagators can be resolved in detail.
Exchange propagators $B(p)$, $p = (p_0, \bop)$, are thus written
\begin{equation}\label{eq:sector-discretisation}
  B(p) = \sum_{\mathrm{Segments}\ S} B^S(p_0)\ \chi_S(\bop)
,
\end{equation}
with $\chi_S$ the characteristic function of segment $S$, and $B^S$
describes frequency dependence inside segment $S$.
By symmetry one can
restrict the discretization procedure to one eighth of the BZ, e.g. to momenta
$\{ (x, y): 0 \le y \le x \le \pi \}$, see Fig.~\ref{fig:sector-discretisation}.
The actual choice of segments can be individually adapted to the type of exchange propagators
for optimal resolution. A quadratic spacing in the radial direction allows a
detailed study
of the $D_{22}$ propagator. 
Linear spacing is used for all other exchange propagators.

The frequency dependence of exchange propagators inside each momentum segment can be 
taken into account in several ways.
The simplest possibility is to assume constant functions 
$B^S(p_0) \equiv B^S( p_0 = 0 )$. 
The restriction to zero frequency has been widely made in the past
and focuses on the frequency value of most singular behavior.

The suitability of more elaborate parametrizations which account for
the decay of exchange propagators in the frequency variable has been examined
in Ref.~\onlinecite{HGS2011}.
Motivated from the small/large frequency asymptotics of RPA resummations
it is natural to employ the specific functional form of a Lorentz-like distribution,
i.e.\ a rational function 
$\om \mapsto (a^2_\bop + ib_\bop\ \om + c^2_\bop\ \om^2)^{-1}$
with momentum-dependent parameters $a_\bop, b_\bop, c_\bop \in \RR$.
These three parameters can then be determined from the flow equation
e.g. from the values of the exchange propagators and their first and second order frequency derivatives
 at zero transfer frequency.
However, 
it seems that only small-frequency behavior is captured well
by this procedure, which is insufficient for precise calculations.
 
Here we do not make an assumption on the functional dependence on the 
transfer frequency variable, but instead discretize it on a frequency grid 
$( \om_i )_{i = 1}^n$ with logarithmic spacing and
include $\om_0 = 0$. 
This choice allows us to study in detail the
small frequency region where, especially at low scales $\Om$, exchange propagators 
can vary strongly. By symmetry it suffices to take into account only
non-negative frequency values.

The discretization of self-energy is specified in the respective sections.

%%%%%%%%%%%%%%%%%%%%%%%%%%%%%%%%%%%%%%%%%%%%%%%%%%%%%%%%%%%%%%%%%%%%%%%%%%%%%%%%%%%%%%%%%%%%%%%%%%%%%%%%%%%%%%%%%%%%%%%
\section{Fermi surface flow}
\label{sec:FSflow}

In this section, we discuss the detailed setup and the results of a flow in which
the frequency-independent part of the self-energy is taken into account.
This allows us to determine the RG flow of the Fermi line
and the change in the single-electron excitation spectrum due to electronic interactions.
We choose the self-energy to have the same symmetries as the Hamiltonian, 
and in particular, to be invariant under discrete lattice rotations.
The breaking of this symmetry and the corresponding nematic phase transitions
have been studied in Refs.\ 
\onlinecite{MetznerRoheAndergassen2003,Metzner2003_Nematicity,Metzner2006_Nematicity,HusemannMetzner2012}.
An asymmetry of $x$ and $y$-direction could be introduced in our study
without any technical difficulties. 
Here we have chosen a symmetric self-energy because 
the Pomeranchuk instability indicating a nematic phase was previously not found 
to dominate the other instabilities of the symmetric state in the RG flow, 
hence we do not expect it to set the scale where the symmetric phase ends.
At the quantum critical point itself, the growth of the interactions
gets suppressed altogether. 
Clearly, the nematic tendency must be taken into account below critical scales, 
e.g.\ in its interplay with the $d$-wave correlations, but this is outside
the range of validity of our flow.

%%%%%%%%%%%%%%%%%%%%%%%%%%%%%%%%%%%%%%%%%%%%%%%%%%%%%%%%%%%%%%%%%%%%%%%%%%%%%%%%%%%%%%%%%%%%%%%%%%%%%%%%%%%%%%
\subsection{Flow equations for hopping correction parameters}

The frequency-independent part of the self-energy satisfies

\begin{align}\label{eq:Sigma0flow}
  \dot\Sigma(p_0 = 0, \bop)
  =&
  \undemi ( -2U + K_{11}(0) )\ \int \intd l\ s(l)
\nonumber\\
  &+ \undemi \int \intd l\ s( l_0, \bol + \bop ) 
     \Big( B_{11}(l) + B_{22}(l)\ f_2^2( \frac{\bol}{2} + \bop ) \Big)
,
\end{align}
where the interaction vertex enters via the linear combination of exchange propagators
$B_{11} = \undemi( 2D - 3M - K )_{11}$ and $B_{22} = D_{22}$,
and $s(l) = s_\Om(l)$ denotes the single-scale propagator. 
Here we use the approximation of
a frequency-independent interaction vertex and set all frequency variables to 0. 
By symmetry, $\Sigma(p_0 = 0, \bop) \in \RR$.

Note that $\Sigma(0, \bop)$ satisfies the symmetries \eqref{eq:Sgm_symmetries} 
of the free dispersion relation.
This is why we parametrize the frequency-independent self-energy
as a sum of corrections to hopping terms 
\begin{equation}\label{eq:ReSgm_parametrisation}
  \Sigma( 0,  \bop )
  =
  \sum_{i \ge 0} c_i\ h_i( \bop ),
  \qquad c_i \in \RR
,
\end{equation}
with functions $h_i$ chosen from an ONS
w.r.t.  
$\skp{f}{g} = \int \intdbar\bop\ \overline{f(\bop)}g(\bop)$.
The first elements of this system are
\begin{equation}\begin{split}
  h_0( x, y ) &= 1,
  \\
  h_1( x, y ) &= \cos x + \cos y,
  \\
  h_2( x, y ) &= 2  \cos x \cos y, 
  \\
  h_3( x, y ) &= \cos 2x + \cos 2y, 
  \\
  h_4( x, y ) &= \sqrt{2} ( \cos 2x \cos y + \cos x \cos 2y ).
\end{split}\end{equation}
We set up a flow at constant particle density, which fixes the
coefficient $c_0$; details are given below.
The remaining coefficients $(c_i)_{i \ge 1}$ can be determined in different ways,
which in general will give different results.
The first is a determination of the $c_i$ from a global average in the
$L^2$ sense:
From the ansatz \eqref{eq:ReSgm_parametrisation} the flow of 
coefficients $c_i$, $i \ge 1$, is uniquely given by orthogonal projection
of $\dot\Sigma(0, \bop)$. 
Orthogonal projection 
$\dot c_i = \skp{ h_i}{ \dot\Sigma(0, \cdot) }$ 
yields
\begin{equation}\label{eq:cidot}\begin{split}
  \dot c_i 
  &=
  - \fracun{2} \int \intd l\ s(l)\ R_i( \bol ),	\quad ( i \ge 1 ),
  \\
  R_i( \bol )
  &=
  \int \intdbar \bop\ 
  	h_i( \bol - \bop )\ 
	\left( B_{11}(0, \bop ) + B_{22}(0, \bop )\ f_2^2( \bol - \frac{\bop}{2} ) \right)
.
\end{split}\end{equation}
Here,
the dependence of $R_i(\bol)$ on $\bol$ 
can be extracted from the momentum integral
again by trigonometric identities. %sum formulas for the cosine function. 
We have furthermore 
$R_i( \hat T \bol ) = R_i ( \bol )\ \forall \hat T \in \TT,\ i \ge 1$, 
since the functions $h_i$ have this property.

Another method to fix the $c_i$ is by local information at special points in 
momentum space,
such as a Taylor expansion of $\dot\Sigma(0, \bop)$ around the Van Hove points.
Comparison with the proposed parametrization then results in a linear system 
that can be solved for the $\dot c_i$.
Whereas the above orthogonal projection method allows to determine
the $c_i$ uniquely,
the local expansion of $\Sigma(0, \bop)$ in general involves all hopping modes
that appear in \eqref{eq:ReSgm_parametrisation}.
Therefore, a finite linear system using local information
extracts an approximation to (finitely many) coefficients $c_i$.
Focusing on the saddle point region $\bop = (0, \pi)$ is natural since the Van Hove singularity substantially
drives RG flows. In principle, the Hessian $(\partial^2 \dot\Sigma)(0, (0, \pi))$ provides two
equations for fixing $\dot c_1$ and $\dot c_2$. 
However, using only two expansion parameters, we observe a discrepancy to 
the above orthogonal projection method. 
In fact, this highly local procedure completely disregards self-energy flow
for momenta away from the saddle point region,
which, though, significantly
feed in via the constant particle density condition,
and can lead to unstable behavior.

This is why we supplement the linear system by two further equations 
such that 
focus on the saddle point region is lifted.
We consider the (non-degenerate) system
\begin{equation}\label{eq:Sgmdot_linearsystem}
  \begin{array}{l c r r r r r}
    \left( (-\partial_x^2 + \partial_y^2)\ \dot{\Sigma} \right) (0, (0, \pi))
    &=& 
    & 2\dot c_1 & & &-6\sqrt{2}\dot c_4
\\
    \left( (\partial_x^2 + \partial_y^2)\ \dot{\Sigma} \right) (0, (0, \pi))
    &=& 
    & & 4\dot c_2 & -8\dot c_3 &
\\
    \dot\Sigma (0, (0, 0)) -  \dot\Sigma (0, (0, \pi))
    &=& 
    & 2\dot c_1 & +4\dot c_2 & & +2\sqrt{2}\dot c_4 
\\
    \dot\Sigma (0, (\pi, \pi)) -  \dot\Sigma (0, (0, \pi))
    &=& 
    & -2\dot c_1 & +4\dot c_2 & & -2\sqrt{2}\dot c_4 
  \end{array}
.
\end{equation}
Below, we compare the results obtained from the two methods.

\begin{figure}
 \includegraphics{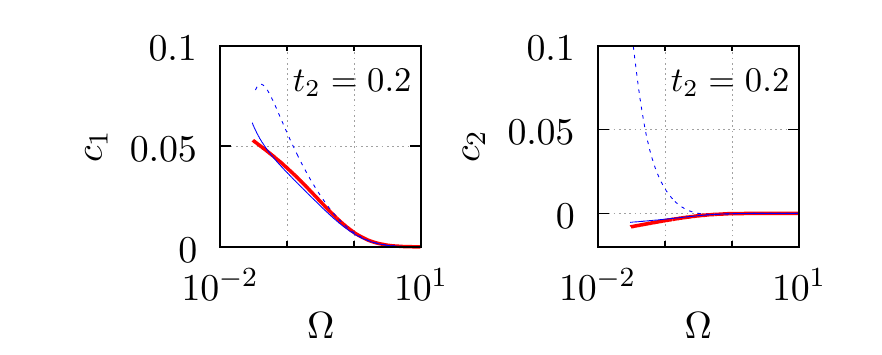}

 \includegraphics{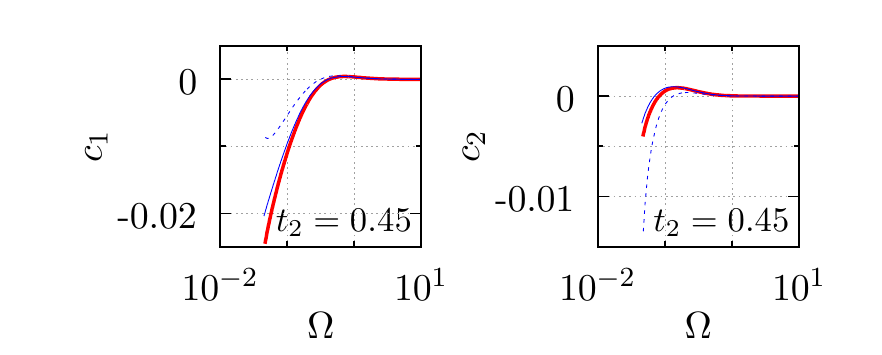}

 \includegraphics{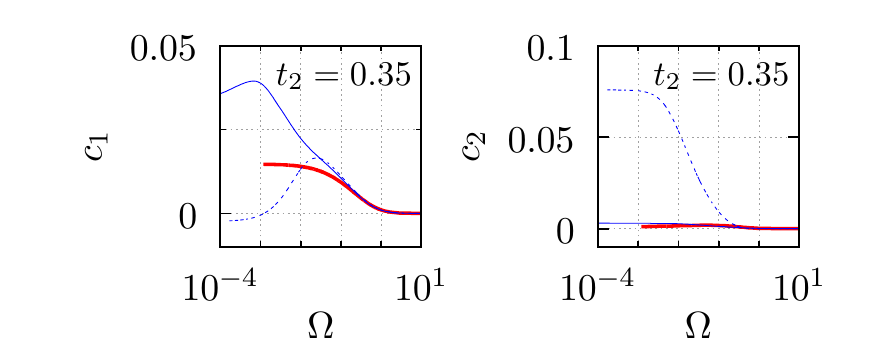}

  \caption{Flow of hopping corrections in the orthogonal projection
scheme(thick lines) vs. the local expansion scheme(thin lines).
The dashed line corresponds to considering only the linear system from the 
second order derivative of $\Sigma$ at the saddle point,
the solid curve results from extending this system by 2 further equations as described in the text. 
Both schemes agree well except in a small parameter region about $t_2/t_1 = 0.35$,
where the local expansion method misses the flow to interacting VHF.
All energy scales are given in units of $t_1$.}
  \label{fig:compare_FT}
\end{figure}

%%%%%%%%%%%%%%%%%%%%%%%%%%%%%%%%%%%%%%%%%%%%%%%%%%%%%%%%%%%%%%%%%%%%%%%%%%%%%%%%%%%%%%%%%%%%%%%%%%%%%%%%%%%%%%
\subsection{Fixing the particle density}
In an RG setup with flowing self-energy the Fermi line can change shape and level during flow. 
Consequently, the particle number will in general become scale-dependent.
We consider the flow at constant particle density
\begin{align}
  n_\Om 
  :=
  \int \intd k\ g_\Om(k)
  \equiv
  n
  ,
\end{align}
i.e. we adjust the self-energy zero mode $c_0$ such that 
$ \dot n_\Om = 0 $.
From the propagator relation $\dot g = s - g^2\dot\Sigma$, the condition of constant 
particle density $\dot n = 0$ reads
\begin{equation}\label{eq:c0dot}
  0 
  =
  \int \intd p\  
    \frac{(ip_0 - \eps_\bop)\ \dot\chi(p_0)}{ \Big( ip_0 - \eps_\bop + \chi(p_0)\Sigma(p) \Big)^2 } 
  + \int \intd p\  
    \frac{ \chi^2(p_0) \dot\Sigma(p) }{ \Big( ip_0 - \eps_\bop + \chi(p_0)\Sigma(p) \Big)^2 } 
\end{equation}
and can easily be solved for %the parameter 
$\dot c_0$,
since $\dot c_0$ enters only linearly.

We choose the particle density such that 
the Fermi line of the system at the stopping scale 
with effective dispersion 
$\eps_\bop^\textnormal{eff} := \eps_\bop - \Sigma_{\Om_*}(0, \bop)$
contains the Van Hove points.
We loosely refer to this situation as ``interacting Van Hove filling'',
although full removal of the regularization at $T = 0$ requires 
a non-symmetric vertex parametrization.

\begin{figure}
 \includegraphics{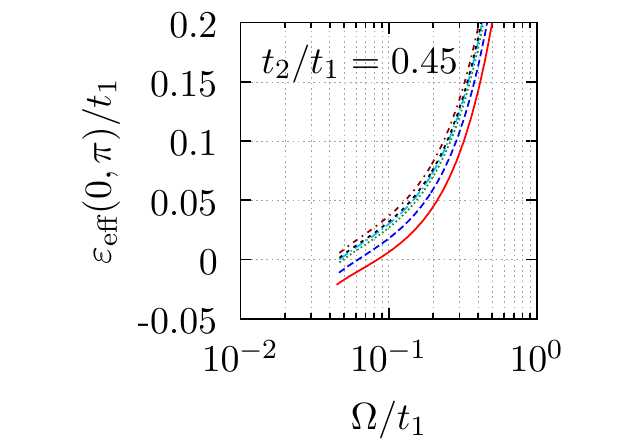}
 % \beginpgfgraphicnamed{fig_mueff} 
 %  \input{abcFnew_moderate-mlarge_t20_45_-mueff}
 % \endpgfgraphicnamed

  \caption{Scale dependence of the effective saddle-point energy level
     $\eps^\Om_\textnormal{eff}(0, \pi) = \eps(0, \pi) - \Sigma_\Om(0, (0, \pi)) = -\mu - c_0 + 2c_2$
     in the calculation considering $c_1$ and $c_2$.
     The flow is calculated at hopping ratio $t_2/t_1 = 0.45$ and for different values of the chemical potential 
     $\mu/t_1 = 0,\ -0.0052, -0.0097, -0.011(*), -0.012, -0.015$ (bottom-up).
     The choice $(*)$ leads to $\eps^{\Om_*}_\textnormal{eff}(0, \pi) = 0$.
   }
   \label{fig:fs-flow_mueff}
\end{figure}

The adjustment of the particle number $n$ is accomplished
via proper choice of the chemical potential $\mu$,
\ie several RG flows with different $n$ need to be calculated,
from which the correct one is selected.
We choose the value of the chemical potential $\mu$ such that
the free system has filling $n$
and do not modify $\mu$ during flow,
as this would leave the framework of 1PI flow equations.
The self-energy degree of freedom corresponding to the chemical potential
is the zero mode $c_0$.
Its initial value is chosen
such that the interacting system at initial scale $\Om_0$ has filling $n$.
This is the appropriate initial condition in the flow at fixed particle density.
Figure~\ref{fig:fs-flow_mueff} shows the flow of the effective saddle point level for different
choices of $\mu$ close to interacting Van Hove filling.

%%%%%%%%%%%%%%%%%%%%%%%%%%%%%%%%%%%%%%%%%%%%%%%%%%%%%%%%%%%%%%%%%%%%%%%%%%%%%%%%%%%%%%%%%%%%%%%%%%%%%%%%%%%%%%
\subsection{Numerical implementation and results at Van Hove filling}

In the numerical implementation of the flow equations,
frequency-momentum loop integrals are evaluated as follows:
By our choice of the regulator function \eqref{eq:Om_regulator},
and because we take only $\Sigma_\Om(0, \bop)$ into account in this section,
the frequency integral can be done analytically using contour techniques. 
To do this for the loop integrands $L_\Om$, which contain
$[ ip_0 - \eps_\bop + \chi_\Om(p_0)\ \Sigma(0, \bop)]^{-1}$,
the
zeros of polynomials of third degree have to be calculated. The remaining momentum integral is
then evaluated numerically.
All calculations have been performed for interaction parameter $U = 3t_1$. 
Exchange propagators are discretized in momentum space by 
60 segments per eighth of the BZ, see Fig.~\ref{fig:sector-discretisation}.

We implement the flow at constant particle density, chosen to be interacting VHF.
This condition fixes the flow of the 
coefficient $c_0$ and thereby essentially determines the flowing
effective saddle-point level
$\eps_{\textnormal{eff}}^\Om(0, \pi) = \eps(0, \pi) - \Sigma_\Om(0, (0, \pi))$
which is of central importance:
it determines the distance of the Fermi level from
the Van Hove points
and thus triggers
a strong enhancement of the flow of the interaction vertex.

During each RG step, scale derivatives are determined in the following order:
First, $(\dot c_i)_{i \ge 1}$ are calculated from eq.~\eqref{eq:cidot};
second, eq.~\eqref{eq:c0dot} provides $\dot c_0$; finally, the scale derivative of
the interaction vertex is computed from eqns.~\eqref{eq:Bmmdot}.

\begin{figure}
  \hspace{-2cm}
  \includegraphics{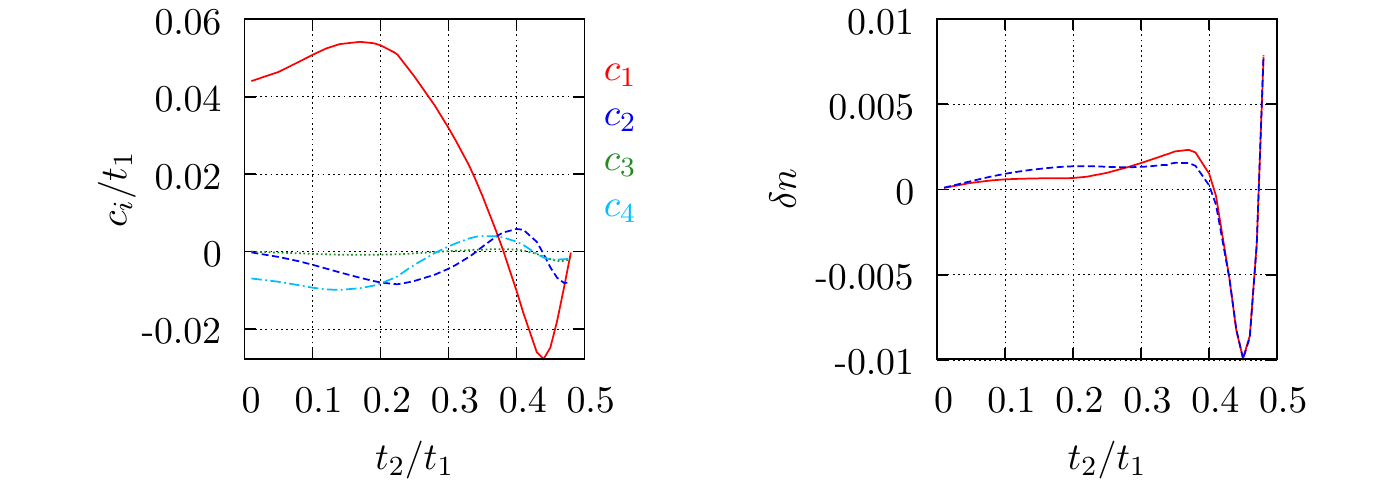}
   \caption{
    Left: Correction parameters $c_1$ to $c_4$ at the stopping scale,
    displayed by the solid, dashed, dotted, dash-dotted line, respectively.
    Right: Particle density corresponding to interacting VHF
    in the orthogonal projection scheme with $2$ (solid)
    and $4$ (dashed) hopping correction terms.
    The plot shows the difference to free VHF.}
    \label{fig:drho_noFreq-wSig}
\end{figure}

\bigskip
The hopping corrections calculated from the RG flow 
remain very small compared with the initial parameters (thick line in Fig.~\ref{fig:compare_FT}).
During the flow, $t_1$ and $t_2$ get a correction of at most a few percent.
Figure~\ref{fig:drho_noFreq-wSig} shows
the value of correction parameters $c_1, \dotsc, c_4$ at the stopping scale
and
the corresponding fine-tuned particle density for interacting VHF.
The dominant hopping correction from the stationary self-energy 
is between nearest neighbors,
as given by the coefficient $c_1$. 
When altering the free FS geometry,
$c_1$ changes sign at $t_2/t_1 \approx 0.38$.
As a consequence we observe that the 
effective interacting FS, determined from $\eps_\bop^\textnormal{eff}$, 
is less curved than the non-interacting one for $t_2/t_1 \lesssim 0.4$,
for larger ratios it is more curved;
this is consistent with previous results.\cite{HonerkampSalmhofer_Umklapp, Katanin_SigmaFM}
The adaptation of the FS patching scheme to the moving FS
in combination with a certain assumption on the radial momentum dependences
of the interaction vertex
similarly finds a
considerable impact of the self-energy zero-mode
whereas the change to hopping parameters $(t_2/t_1)_{\textrm{eff}}$ 
is not too strong.\cite{Katanin_SigmaFM}
Ref.~\onlinecite{HalbothMetzner97} gives the FS deformation
as a function of the filling $n$ at $t_2 = 0$,
calculated from second-order perturbation theory.

\begin{figure}
  \hspace{-1cm} 
  \includegraphics{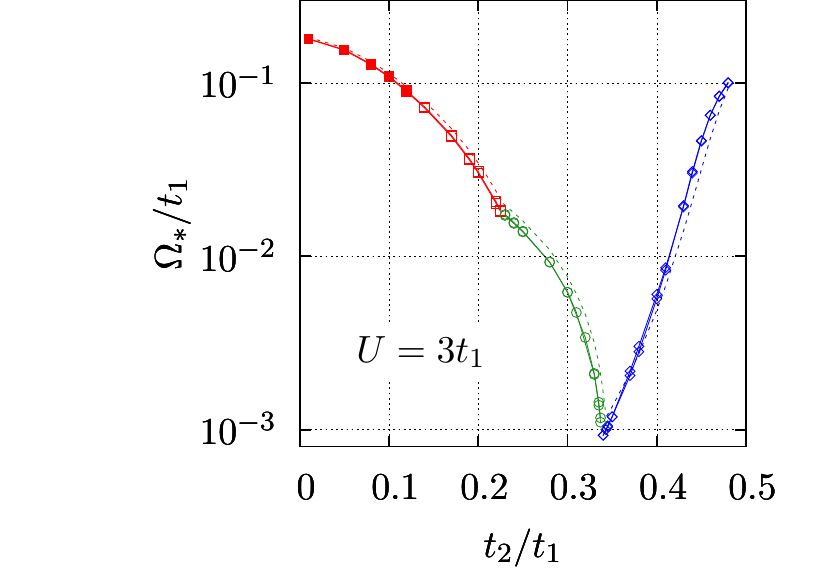}

  \caption{Influence of the moving Fermi line on the flow of the interaction vertex for systems at VHF. 
    Tracing $\Sigma(0, \bop)$
     leads to a small change in the stopping scale
    compared to neglecting self-energy (dashed line). 
    We have calculated self-energy flows with $2$ as well as with $4$ hopping correction terms;
    this no more produces a difference in the stopping scale, the two solid lines are practically
    indistinguishable. Symbols indicate most dominant ordering tendency: 
    commensurate AFM (filled square), incommensurate AFM (open square), $d$-SC (circle),
     FM (diamond).}
   \label{fig:Omcrit_noFreq-wSig}
\end{figure}

The fact that $\Sigma(0, \bop)$ remains small during the flow results in a minor
modification of the structure of the interaction vertex when including the FS flow.
In Fig.~\ref{fig:Omcrit_noFreq-wSig} we compare the stopping scale in the setup
disregarding self-energy (dashed line) to the calculation done here taking account of self-energy
corrections through coefficients $c_1$ to $c_2$ and $c_1$ to $c_4$ (two solid lines, 
almost identical).
Consideration of the FS flow only slightly alters the stopping scale.
Moreover, parameter regions corresponding to different ordering tendencies 
practically do not change.
This indicates that the
coefficients $c_3$, $c_4$ have almost no influence 
on the interaction vertex flow,
hence that a parametrization 
of $\Sigma(0, \bop)$ by the first hopping terms suffices.

\medskip
It is an interesting test for the parametrization
whether the local expansion method \eqref{eq:Sgmdot_linearsystem}
can produce a good approximation to the
flow of hopping corrections. 
Concerning the coefficients $c_1$ and $c_2$,
 we find that this is indeed true for many parameter values.
It is essential here not to focus on the saddle point region only when setting up
flow equations.
In this check
we employ the particle density value for interacting VHF identified within the orthogonal
projection scheme.
The two last equations of the system \eqref{eq:Sgmdot_linearsystem}
then provide,
compared to consideration of $(\partial_\bop^2\Sigma)(0, (0, \pi))$ only,
corrections
that bring $\dot c_1$ and $\dot c_2$ close to the %(exact)
scale derivative
obtained by orthogonal projection, compare the thin solid and dashed blue line in Fig.~\ref{fig:compare_FT}.
The same is  expected for the parameters $c_3$, $c_4$
after appropriate extension of the linear system.

Only
in the region $t_2 \approx 0.35 t_1$ 
of competing pairing and ferromagnetic ordering tendencies,
differing flows for the first two hopping corrections are found.
This is mainly due to the fact that deviations in $\dot c_1$ and $\dot c_2$ %,
result here,
via the constant particle density condition \eqref{eq:c0dot}, in a significantly deviating flow
of $c_0$ such that interacting VHF is 
reached in the orthogonal projection scheme
whereas it is not reached in the local scheme.
This difference has a strong influence to the flow of the interaction vertex
and leads to qualitatively different behavior.
Given the strong influence of the effective saddle-point level,
the flow is sensitive to 
which information is used to fix the particle density via \eqref{eq:c0dot}:
determination of the particle density from an only
locally known $\Sigma(0, \bop)$ can be unreliable,
and the orthogonal projection method should be preferred.

Away from this parameter region and at interacting VHF,
a parametrization of $\Sigma(0, \bop)$ by the first few hopping terms provides consistent results.
We observe that the FS flow has almost no impact on the interaction
vertex.

%%%%%%%%%%%%%%%%%%%%%%%%%%%%%%%%%%%%%%%%%%%%%%%%%%%%%%%%%%%%%%%%%%%%%%%%%%%%%%%%%%%%%%%%%%%%%%%%%%%%%%%%%%%%%%%%%%%%%%%
\section{Frequency-dependent self-energy}\label{sec:ImSigma}

We now include the frequency dependence of the self-energy
and its influence on the flow of effective interactions,
by solving the RG equations where the frequency- and momentum-dependent 
self-energy appears in the full propagators. 
Here we neglect the real part of the self-energy
since its frequency-independent value was found to remain small during the symmetric flow,
see Sec.~\ref{sec:FSflow}.

%%%%%%%%%%%%%%%%%%%%%%%%%%%%%%%%%%%%%%%%%%%%%%%%%%%%%%%%%%%%%%%%%%%%%%%%%%%%%%%%%%%%%%%%%%%%%%%%%%%%%%%%%%%%%
\subsection{RG setup}

Within the stationary vertex approximation, the flow equation \eqref{eq:Sigma_v_flow}
generates a frequency-independent self-energy. 
A non-trivial flow equation for the frequency-dependent self-energy can be obtained though
from the stationary interaction vertex
by inserting its scale-integrated flow equation into the one of
the two-point function.\cite{Honerkamp_ScatRate, HonerkampSalmhofer_QuasiparticleWeight, KataninKampf2004, Katanin-TwoLoop}
The parametrization of the frequency dependence of the interaction vertex
enables us here to calculate the full frequency dependence
of the self-energy,
beyond the above approximation.
To this end the set of RG equations for the frequency-dependent interaction vertex\cite{HGS2011}
is extended by a self-energy flow equation.
Furthermore, the full self-energy feed-back to the flow is accounted for
by keeping full propagators on the right-hand side of flow equations.

The most singular frequency dependence of self-energy is expected for its imaginary part
and at small frequencies.
Analysis of $\Sigma(p)$ in second order perturbation theory 
for a Hubbard-like system at VHF and with a square Fermi surface shows\cite{FeldmanSalmhofer2008b}
a logarithmic singularity in the first frequency derivative for momentum $\bop = (0, \pi)$:
At zero temperature and for frequencies $0 < \om < \undemi$,
\begin{equation}\label{eq:Z_ptb_asymptotics}
( \partial_{\om}\rIm\Sigma^ {(2)} ) (\om , \bop = (0, \pi)) 
= 
C_1 \abs{\ln \om}^2 + C_2 \abs{\ln \om} + \cO( \om^0 )
\end{equation}
with $C_1, C_2 > 0$.
It is also shown in Ref.~\onlinecite{FeldmanSalmhofer2008b} that
$\nabla\Sigma$ remains small to all orders in the coupling, even at VHF.
Therefore, the growth \eqref{eq:Z_ptb_asymptotics} can have a drastic effect.

\begin{figure}
  \hspace{-3em}
 \includegraphics{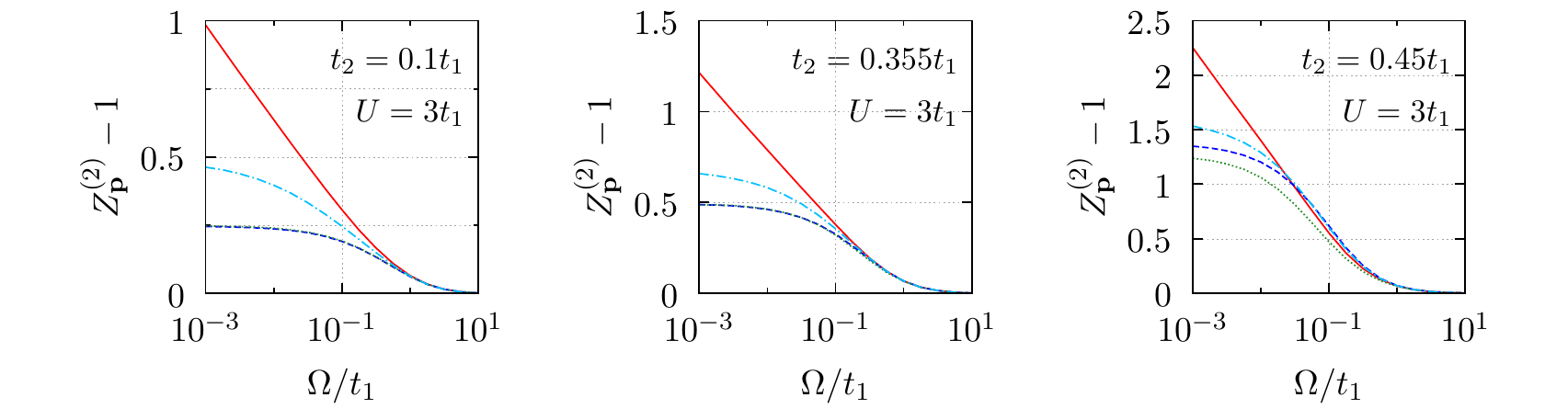}
  \caption{Second order perturbation theory for the $Z_\bop$ factor at VHF 
    and momenta $\bop = $ $(0, \pi)$, $(x_0, x_0)$, $(0, 0)$,
    $(\pi, \pi)$ (solid, dash-dot, dashed, dotted lines, respectively;
    $\bop = (0, 0), (\pi, \pi)$ almost coincide in the first two plots). 
    The abscissa $x_0$ provides the intersection point of FS with the BZ diagonal.
}
    \label{fig:dSigma_ptb}
\end{figure}

Perturbation theory for the Hubbard system with $\Om$ regularization,
at VHF and for a curved Fermi surface, $ 0 < t_2/t_1 < \undemi$,
yields a similar picture. 
Numerical calculation of the $Z$ factor,
\ie the inverse quasi-particle weight,
\begin{equation}
  Z_\bop = 1 + (\partial_\om \rIm\Sigma)(\om = 0, \bop)
\end{equation}
to second order in $U$
yields a logarithmic divergence (with power $1$) in the scale parameter $\Om$
for $\bop = (0, \pi)$ and indicates regular behavior for all other momenta, see Fig.~\ref{fig:dSigma_ptb}.

%Therefore, 
For a detailed study of effects from the frequency-dependent self-energy
we consider the flow equation for its imaginary part
\begin{align}\label{eq:ImSgmdot}
   \rIm \dot\Sigma(p)
    =
    -\undemi \int\intd l\
    \bigg[
       &\rIm s(l+p)  
      \left( (\rRe D + \undemi K + \frac{3}{2} M)_{11}(l) 
             + \rRe D_{22}(l) f_2^2( \frac{\bol}{2} + \bop ) \right)
     \nonumber\\
      %
      %&
      +
     &\rRe s(l+p)  
      \left( \rIm D_{11}(l) + \rIm D_{22}(l) f_2^2( \frac{\bol}{2} + \bop ) \right)
    \bigg]
.
\end{align}
The propagator including the imaginary part of the self-energy reads
$ g(p) =   \chi(\om)\ \big( i\om - \eps_{\bop} + i\chi(\om) \rIm\Sigma(p)\big)^{-1}$
and enters the vertex flow equation \eqref{eq:Bmmdot} via
\begin{align}\label{eq:LwithImSgm}
&
  L(p_1, p_2)
=
 \diff{\Om} \Big( g(p_1) g(p_2) 
 \Big)
\nonumber\\
&=
   g(p_1) g(p_2)
   \left(
     \frac{\dot\chi(\om_1)}{\chi(\om_1)} + \frac{\dot\chi(\om_2)}{\chi(\om_2)}
     - \frac{i \diff{\Om} \big( \chi(\om_1)\rIm\Sigma(p_1) \big) }{ i\om_1 - \eps_{\bop_1} + i\chi(\om_1) \rIm\Sigma(p_1)}\
     - \frac{i \diff{\Om} \big( \chi(\om_2)\rIm\Sigma(p_2) \big) }{ i\om_2 - \eps_{\bop_2} + i\chi(\om_2) \rIm\Sigma(p_2)}
\right)
.
\end{align}

Since $\rIm\Sigma$ is expected to show non-regular behavior
only in the small frequency region, 
a linear frequency parametrization 
$\rIm\Sigma(\om, \bop) \approx (Z_\bop - 1)\ \om$
appears natural. 
However, we find that this linear approximation leads to
a strong artificial suppression of the interaction vertex during the flow, 
as compared
to calculations using a discretization in frequency space.
This strong suppression is due to the contributions from 
intermediate to large frequencies.
Here the linear frequency approximation highly overestimates 
$\rIm\Sigma(\om, \bop)$, 
which, in fact, decays to $0$ as $\abs{\om} \to \infty$.

%%%%%%%%%%%%%%%%%%%%%%%%%%%%%%%%%%%%%%%%%%%%%%%%%%%%%%%%%%%%%%%%%%%%%%%%%%%%%%%%%%%%%%%%%%%%%%%%%%%%%%%%%%%%%
\subsection{Numerical implementation}
 
The flow equations \eqref{eq:Bmmdot} and \eqref{eq:ImSgmdot},
including \eqref{eq:LwithImSgm} and a similar equation 
for the single-scale propagator $s(p)$,
are studied numerically.
For this purpose, exchange propagators are discretized 
in frequency and momentum arguments  by
16 momentum segments per eighth of BZ,
as described in eq.~\eqref{eq:sector-discretisation}.
Inside each momentum segment 
we use a logarithmic grid of length $20..25$
for resolution of transfer frequencies $10^{-6} \le p_0/t_1 \le 10^3$.
The smallest non-zero frequency value here is chosen depending on the expected stopping scale.
Furthermore, the frequency value $p_0 = 0$ is included in the grid.
The function $\om \mapsto \rIm\Sigma(\om, \bop)$ is discretized on a similar
logarithmic frequency grid for a fixed set of momenta $\bop$.
The most singular behavior of $\rIm\Sigma$ is expected at the momentum $\bop = (0, \pi)$. 
Feed-back of $\rIm\Sigma$ to the flow is done by spline interpolation of discrete data.
All calculations are done for the interaction parameter value $U = 3t_1$.

Although the parametrization of $\rIm\Sigma$ as a function of only one
frequency-momentum argument is straightforward and easily accomplished in a 
discretization procedure with a comparably small number of parameters,
the numerical evaluation of flow equations 
involving a frequency- and momentum-dependent self-energy
turns out to be highly time-consuming.
This is due to 
the presence of the full propagator $G = ( C^{-1} + \Sigma )^{-1}$
instead of the bare propagator $C$ on the right-hand side.
For the evaluation of the right-hand side integrals, 
we have set up an analytic loop frequency integration with contour techniques
(\ie after fit of the discrete frequency data to sums of Lorentzians)
or with piecewise analytic integration of spline data\cite{SchmidtEnss}
but found both to be impractical due to the high powers of the frequency variable
in the resulting rational frequency integrand.
Thus a numerical evaluation of the three nested integrals is necessary. 

In the parameter regions
$\abs{t_2/t_1} < 0.2$ and $0.45 < t_2/t_1 < 0.5$
we find stopping scales $\Om_* \gtrsim 10^{-2} t_1$.
Here, 
the full flow including feed-back of the frequency- and momentum-dependent
imaginary self-energy has been numerically calculated.
As the momentum dependence of $\rIm\Sigma(\om, \bop)$ turns out
to vary not too strongly,
a simple momentum set $\{ (0,0),\ (0, \pi),\ (\pi, \pi) \}$
is used for its discretisation.
Results will be shown later in this section, see figs.~\ref{fig:Z-flow}, 
\ref{fig:exampleFlows_wFreq-wSig}.

For parameters $0.2 \lesssim t_2/t_1 \lesssim 0.45$,
stopping scales $\Om_*$ drop below $10^{-2} t_1$.
Because of numerical complexity we then proceed in a different way:
For a fixed momentum value $\boq$ we decompose
\begin{align}\label{eq:G-decomp_ImSgm}
  g(p)
=&
  \frac{ \chi(\om) }{ \Big( i\om - \eps_\bop + i\chi(\om) \rIm\Sigma(\om, \boq) \Big)
    +i\chi(\om) \Big( \rIm\Sigma(p) - \rIm\Sigma(\om, \boq) \Big) }  
\nonumber\\
=&
  \frac{\chi(\om)}{a(p)}  +  \frac{-\Delta(p)}{a(p)\ (a+\Delta)(p)}\ \chi(\om),
\end{align}
with
$ a(p) =  i\om - \eps_\bop + i\chi(\om) \rIm\Sigma(\om, \boq) $
as well as
$ \Delta(p) = i\chi(\om) \Big( \rIm\Sigma(p) - \rIm\Sigma(\om, \boq) \Big)$.
Now
\begin{align}
  \frac{\chi(\om)}{a(p)}
=&
 \frac{ \chi(\om) } {i\Big( \om +\chi(\om) \rIm\Sigma(\om, \boq) \Big) - \eps_\bop } 
=
 \frac{ \chi(\om) } {i r(\om) - \eps_\bop } 
\end{align}
factorizes in a $\bop$ independent term $\chi(\om)$ and 
a term that depends on scale only through 
$r(\om) = \om + \chi_\Om(\om) \rIm\Sigma_\Om(\om, \boq)$.
We insert this decomposition in the right-hand side of flow equations.
For all terms that do not contain the function $\Delta$,
the loop integration can be reduced to the frequency integration
by computing appropriate momentum integrals before the actual RG flow is started,
similar to the treatment in Ref.~\onlinecite{HGS2011}.

To be specific, the flow equations for exchange propagators are of the form
\begin{align}\label{eq:floweq-loopcalc}
  \dot B_{mm}^\Om(p)
=&
  \mp\undemi \int_\RR \intdbar l_0\ \sum_{j_1, j_2} \alpha_{j_1}^\Om(l_0) \alpha_{j_2}^\Om(l_0)\  
\bigg\{
 \dero{\Om} \Big( \chi_\Om(\om_-) \chi_\Om(\om_+) \Big)\ F_{\psi_{j_1}\psi_{j_2}}\Big( r_\Om(\pm\om_-), r_\Om(\om_+), \bop\Big )
\nonumber\\&\quad
 + \chi_\Om(\om_-) \chi_\Om(\om_+) \bigg[
   \dero{\Om} \Big( \chi_\Om(\om_-) \rIm\Sigma_\Om(\om_-, \boq) \Big)\  
      \fracun{i}H_{\psi_{j_1}\psi_{j_2}}\Big( r_\Om(\pm\om_-), r_\Om(\om_+), \bop \Big)
\nonumber\\&\hspace{18ex}
+  \dero{\Om} \Big( \chi_\Om(\om_+) \rIm\Sigma_\Om(\om_+, \boq) \Big)\  
      \fracun{i}H_{\psi_{j_1}\psi_{j_2}}\Big( r_\Om(\om_+), r_\Om(\pm\om_-), \bop \Big)
\bigg] 
\nonumber\\&\quad
+ \int\intdbar\bol\ f[a, \Delta](p, l, \boq)\  \psi_{j_1}(\bop, \bol)\psi_{j_2}(\bop, \bol)
\bigg\}
\end{align}
with
\begin{align}
   F_{\psi_{j_1}\psi_{j_2}}( r_1, r_2, \bop )
=&
   \int\intdbar\bol\ \frac{ \psi_{j_1}(\bop, \bol)\ \psi_{j_2}(\bop, \bol) }
     { (ir_1 - \eps_{\bol - \bop/2})\ (ir_2 - \eps_{\bol + \bop/2}) },
\nonumber\\
   H_{\psi_{j_1}\psi_{j_2}}( r_1, r_2, \bop )
=&
   \int\intdbar\bol\ \frac{ \psi_{j_1}(\bop, \bol)\ \psi_{j_2}(\bop, \bol) }
     { (ir_1 - \eps_{\bol - \bop/2})^2\ (ir_2 - \eps_{\bol + \bop/2}) },
\nonumber
\end{align}
$\om_\mp = l_0 \mp p_0/2$,
and $f[a, \Delta]$ comprises all terms of the decomposition that depend also on $\Delta$.

Here $\sum_j \alpha_j^\Om(l_0)\psi_j(\bop, \bol)$ denotes the vertex feed-back
to the flow as given by equations \eqref{eq:define-explicit-F_m} or \eqref{eq:F1_approx}.
We choose to employ the approximation \eqref{eq:F1_approx} because it reduces
numerical efforts; we have checked that it is of high accuracy in this setup
by comparing specific scale derivatives.
Hence, functions $\psi_j(\bop, \bol) \in \{ 1,\
f_2(\bol - \undemi\bop)f_2(\bol + \undemi\bop),\ f_2(\bol) \}$
are present in the $j$ summations of eq.~\eqref{eq:floweq-loopcalc}.

Momentum integrations $F_{\psi_{j_1}\psi_{j_2}}, H_{\psi_{j_1}\psi_{j_2}}$ can then
be computed beforehand
as they are independent of scale. 
This needs to be done for all discrete momenta $\{\bop\}$ where exchange propagators
are traced and for all combinations $(\psi_{j_1}, \psi_{j_2})$ that occur.
The singularity of these momentum integrals along the ``frequency'' lines
$r_1 = 0$ or $r_2 = 0$ is essential to the flow and needs to be resolved in detail.
The small-frequency asymptotics of the functions $F$, $H$ can be calculated for the case 
$\psi_{j_1} = \psi_{j_2} = 1$ and for zero momentum exchange
from the knowledge about the free Hubbard density of states.
For example, at VHF and for $0 < t_2/t_1 < \undemi$,
\begin{align}
  F_{11}(r_1, r_2, \bzero)
=&
 \int\intd e\ \frac{\cN(e)}{(i r_1 - e)( i r_2 - e)}
\nonumber\\
=&
  \left\{
    \begin{tabular}{l @{\ :\ } l}
 $- C_1 \big[ \sgn r_2\ \ln\abs{r_2} - \sgn r_1\ \ln\abs{r_1} \big] / \big( r_2 - r_1 \big)
  + \cO(r_1^0, r_2^0)$ & $r_1 \neq r_2$\\
  $- C_2 /{\abs{r_1}} + i C_3 \sgn r_1\ \ln\abs{r_1} + \cO(r_1^0)$ & $r_1 = r_2$
    \end{tabular}
  \right.
\end{align}
with $C_1, C_2, C_3 > 0$.
We use a logarithmic grid on both sides of the point $r_1 = 0$ and similarly for $r_2$,
altogether a two-dimensional grid of $240$ x $240$ points.
The minimal distance in the grid to the lines $r_1 = 0$ or $r_2 = 0$ is carefully adjusted
depending on the minimal value of the scale parameter $\Om$ that is needed during flow.
The flow equation for $\rIm\Sigma$ is treated in the same way.

The loop integration for the terms including $\Delta$ remains a 
highly time-consuming computational task
because three nested integrations need to be performed numerically
at every RG step and for each coupling.
However, 
comparison in the parameter region with high stopping scales as well as further tests 
discussed below suggest
that for the choice $\boq = (0, \pi)$
in the decomposition \eqref{eq:G-decomp_ImSgm}, 
\ie for the momentum value with expected strongest feed-in of $\rIm\Sigma$ to the flow, 
these remaining terms containing $\Delta$ are well dominated by the first term. 

Thus, we can approximately calculate the flow by first neglecting these remaining terms. 
Then, as an error estimate, the approximated and full scale derivatives 
at the thus obtained stopping scale can be compared for specific couplings. 
Below, this approximation is discussed further.
Notice that global momentum-independent feed-in of $\rIm\Sigma(\om, (0, \pi))$
still allows the determination of $\rIm\Sigma(\om, \bop)$ at different points $\bop$ in the BZ.

%%%%%%%%%%%%%%%%%%%%%%%%%%%%%%%%%%%%%%%%%%%%%%%%%%%%%%%%%%%%%%%%%%%%%%%%%%%%%%%%%%%%%%%%%%%%%%%%%%%%%%%%%%%%%
\subsection{Results at Van Hove filling}

Here we discuss results of the flow with frequency-dependent self-energy
where the calculations have been performed using the approximation of 
momentum-independent feed-in of $\rIm\Sigma(\om, (0, \pi))$ to the flow,
as described above.
For parameter values where the full flow could be calculated
this approximation turns out to be very close to the true flow,
details about this point are given in the next section.
We compare the results to previous flows disregarding self-energy effects.

\begin{figure}
 \includegraphics{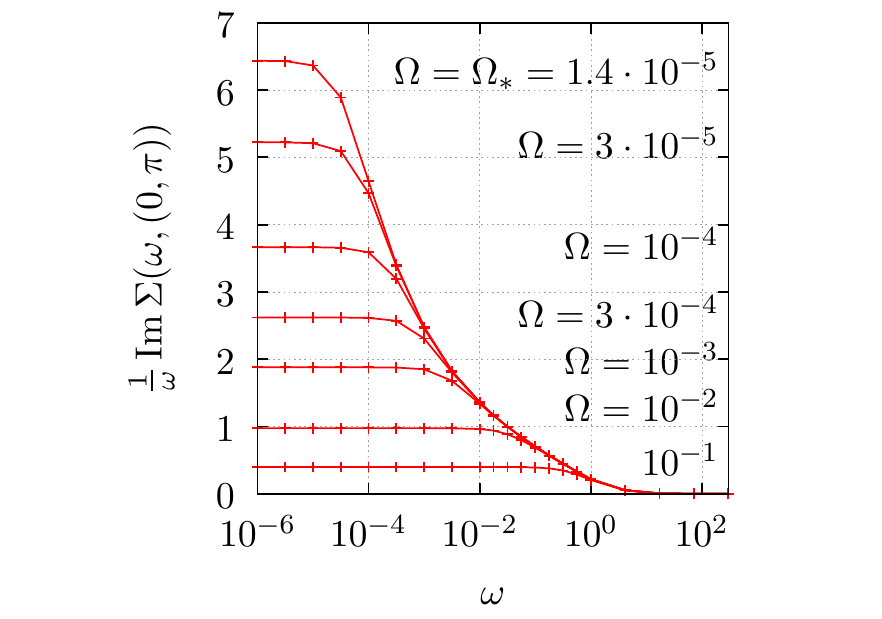}
  \caption{Frequency dependence of $\rIm\Sigma$ for parameter values $t_2 = 0.355t_1$, $U = 3t_1$.
  The function $\om \mapsto \fracun{\om}\rIm\Sigma(\om, (0, \pi))$ is shown for different values
of the scale parameter.
All energies are measured in units of $t_1$.}\label{fig:ImSigma-flow}
\end{figure}

\begin{figure}
  \hspace{-3em}
  \includegraphics{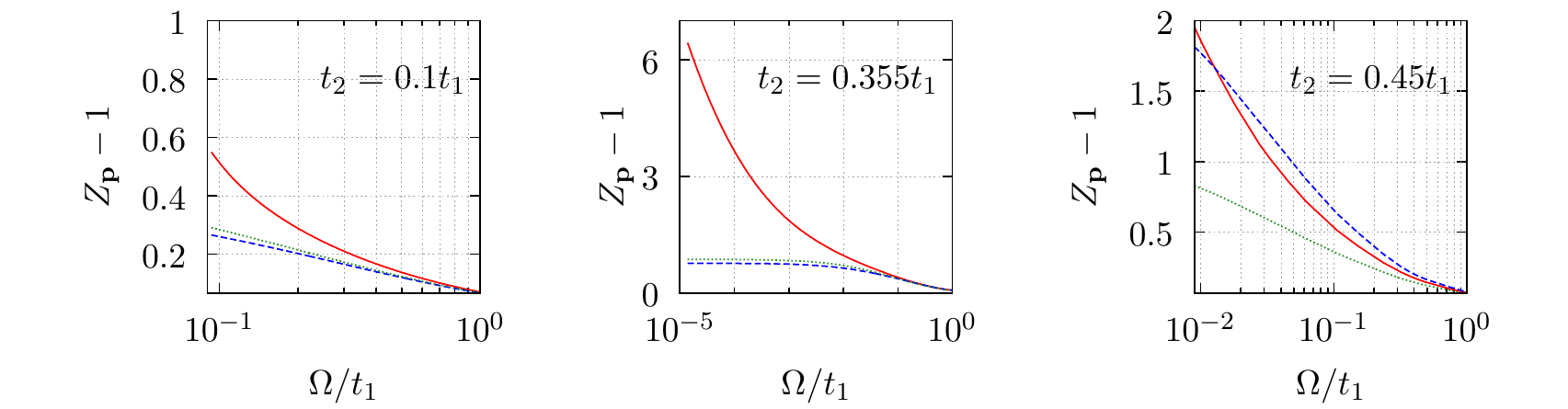}
  \caption{Flow of the $Z_\bop$ factors (inverse quasi-particle weights)
 for $\bop = (0, \pi)$, $(0, 0)$, $(\pi, \pi)$ (solid, dashed, dotted lines, respectively)
 until reach of the varying stopping scales.
 The calculations for $t_2/t_1 = 0.1$ and $0.45$ include feed-back 
 of the full frequency and momentum dependence of $\rIm\Sigma$.
 For $t_2/t_1 = 0.355$ the approximation of the momentum-independent
 feed-back of $\rIm\Sigma(\om, (0, \pi))$ is used.
}\label{fig:Z-flow}
\end{figure}

The frequency-dependent self-energy 
feeds back to the flow via
$g(\om, \bop) = 
\chi(\om)\ 
\big[ i\om \big( 1 +\\
\fracun{\om}\rIm\Sigma(\om, \bop)\ \chi(\om) \big) - \eps_\bop  \big]^{-1}
$.
We show the function 
$\om \mapsto \fracun{\om}\rIm\Sigma(\om, \bop)$ 
in Fig.~\ref{fig:ImSigma-flow}
for an exemplary flow at different scales. 
The limit $\om \to 0$ of this function yields
the momentum-dependent $Z_\bop$ factor. From this plot the region of validity of a linear frequency
dependence of $\rIm\Sigma$ can be seen:
this region is roughly $\abs{\om} \lesssim \Om$.
We have calculated $\rIm\Sigma(\om, \bop)$
for $\bop \in \{ (0, 0), (0, \pi), (\pi, \pi)\}$ and find,
as compared to results of second order perturbation theory,
that $Z_\bop$ factors for all three momenta usually get enhanced in the RG calculation,
cf. figs.~\ref{fig:dSigma_ptb} and \ref{fig:Z-flow}.
In particular, $Z_{(0, \pi)}$ shows a strong divergence as $\Om \to 0$ 
and hence becomes large at small scales.
For $t_2/t_1 = 0.355$, a value of $Z_{(0, \pi)} \approx 7$ 
is reached at the stopping scale, 
whereas $Z_{(0, 0)}$ and $Z_{(\pi, \pi)}$ tend towards finite values in the flow.
For parameter ratios $t_2/t_1$ close to but smaller than $\undemi$, 
the $Z_{(0, 0)}$ factor is driven
by an emerging singularity in the density of states, originating from 
%momentum region$\bop = \bzero$ 
the band minimum instead of the saddle point region.
As a consequence, $Z_{(0, 0)} > Z_{(0, \pi)}$ 
during a substantial part of the flow,
this effect is already present in perturbation theory.

\begin{figure} 
 \includegraphics{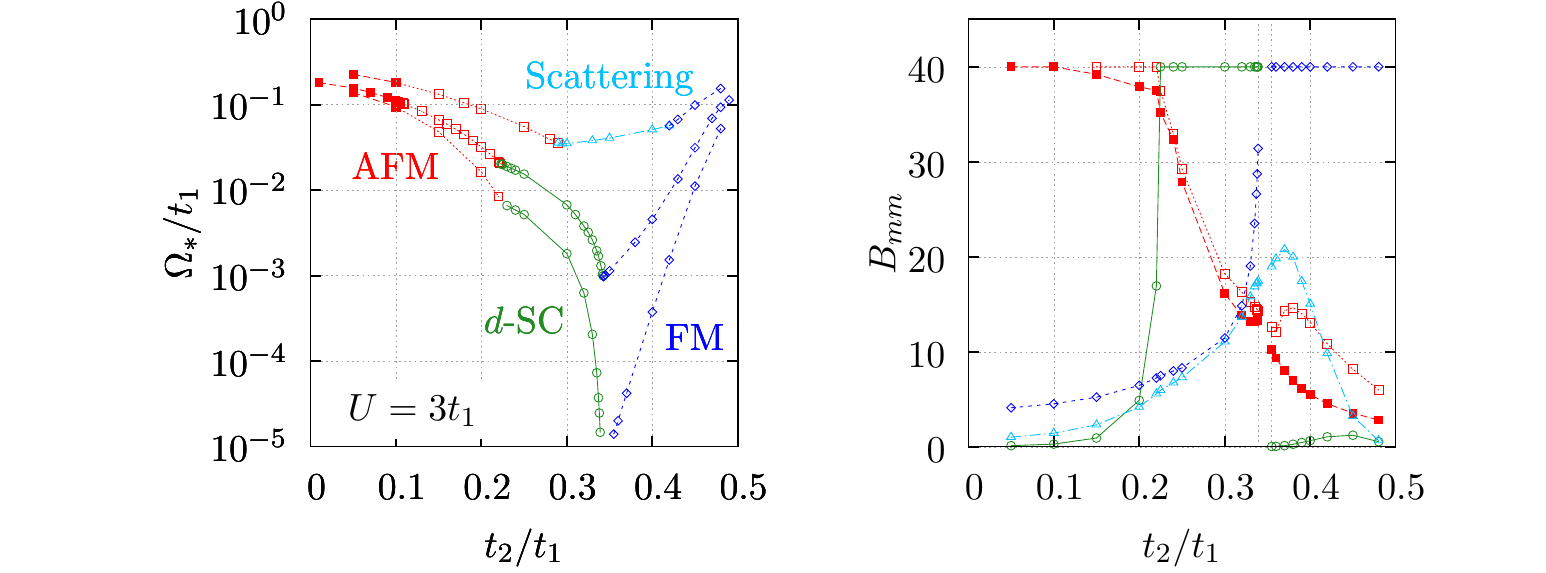}
  \caption{Left: Comparison of stopping scales $\Om_*$ at VHF 
    for different vertex parametrizations.
Top-down: 
(i) frequency-dependent vertex disregarding self-energy,
(ii) frequency-independent vertex disregarding self-energy,
(iii) frequency-dependent vertex including $\rIm\Sigma(\om, (0, \pi))$.
Dominant ordering tendencies: commensurate AFM (dashes, filled square), 
incommensurate AFM (dotted, open square), $d$-SC (solid, circle),
scattering instability (dash-dot, triangle), FM (short dashes, diamond).
Right: Values of largest couplings in the different channels, 
for setup (iii), at the stopping scale.
}
\label{fig:Omcrit_ImSigma-flow}
\end{figure}

The fact that $Z_\bop$ factors become significantly larger than one %$1$ 
leads to a suppression of the flow of the interaction vertex, 
and especially at low scales the divergence of $Z_{(0, \pi)}$ has a strong impact.
Thus, the frequency-dependent self-energy (over-)compensates the effect
of enhanced divergences and rising stopping scales 
from taking the frequency dependence of the interaction vertex into account.\cite{HGS2011}
In Fig.~\ref{fig:Omcrit_ImSigma-flow} we compare stopping scales 
resulting from different vertex parametrizations.
For small ratios $t_2/t_1$ or ratios $t_2/t_1$ near $\undemi$,
the flow of the frequency-dependent interaction vertex including $\rIm\Sigma$
is quite 
 close to a flow
that neglects self-energy and frequency dependences at the same time.
This has already been observed\cite{HonerkampSalmhofer_QuasiparticleWeight}
at $t_2 = 0$,
in a simpler approximation where only a $Z$ factor is kept.
For intermediate hopping ratios, the stopping scale is even further suppressed.
In particular,
in the parameter region 
of competing ferromagnetic
and $d$-wave pairing ordering tendencies the stopping scale drops drastically
to values beyond numerical resolution,
for $0.34 \lesssim t_2/t_1 \lesssim 0.35$ we detect stopping scales
$\Om_* /t_1 < 10^{-5}$. 
This behavior suggests a quantum critical point
separating both ordering regimes, 
which was also proposed in previous studies.\cite{HonerkampSalmhofer_Tflow}

Besides the lowering of the stopping scale we observe two additional differences
to the setup with frequency-dependent vertex disregarding self-energy.\cite{HGS2011}
First, a region of dominant $d$-wave pairing, which had disappeared in the calculation
with frequency-dependent exchange propagators, is recovered.
We see this change as a direct consequence of the small stopping scales
that are obtained now,
which allow effective generation of $d$-SC correlations driven from
the magnetic channel.
Furthermore, the scattering singularity, which had appeared in the study
with frequency-dependent exchange propagators, is weakened and no more
becomes dominant. However, it is still present.
The right-hand plot in Fig.~\ref{fig:Omcrit_ImSigma-flow} shows the values
of the most singular couplings in each channel at the stopping scale. 
The scattering coupling is at most of half the size of the most singular coupling.

Thus, at VHF, the RG flow that takes into account the frequency dependence 
of both the interaction vertex and $\rIm\Sigma$ results in the same 
types of Fermi liquid instabilities as the flow where the vertex is static and 
the self-energy dropped. The parameter regions for dominance
of AF, $d$-SC and FM come out almost identical in these two flows, too.
For interaction parameter $U = 3t_1$, we observe at hopping ratios
$\abs{t_2/t_1} \lesssim 0.2$ a region of dominant AFM
as a consequence of (approximate) nesting.
Upon further increase of the FS curvature, relatively strong AFM
correlations generate in a Kohn-Luttinger-like effect 
an instability in the pairing channel with $d$-wave symmetry.
For ratios $0.35 \lesssim t_2/t_1 \le 0.5$, formation of 
$d$-SC correlations abruptly stops due to emergence 
of a strong ferromagnetic instability, which crucially depends on
the presence of Van Hove points on the FS.

%%%%%%%%%%%%%%%%%%%%%%%%%%%%%%%%%%%%%%%%%%%%%%%%%%%%%%%%%%%%%%%%%%%%%%%%%%%%%%%%%%%%%%%%%%%%%%%%%%%%%%%%%%%%%%%%%%%%%%%
\subsection{Non-Fermi-liquid frequency dependence of the self-energy}\label{non-FL-Sigma}

In the parameter region $t_2/t_1 \approx 0.34$
and for several ratios $U/t_1 = 2.5 \dotsc 3.5$,
the competition between pairing correlations with $d$-wave symmetry
and ferromagnetic correlations decelerates the flow of the interaction vertex.
The regime of validity of the level-two-truncation  is large
and permits to trace the RG flow down to relatively small stopping scales.
Previous works have suggested the existence of a quantum critical point in this parameter region.
\cite{HonerkampSalmhofer_Tflow}
The present findings,
which take into account the frequency dependences of the interaction vertex
and of the self-energy,
support this scenario.

The drop in the stopping scale goes along with a strong reduction of the quasi-particle weight 
at the Van Hove points during the flow.
For momenta away from the saddle point region 
we detect a standard FL frequency dependence
$\Sigma(\om, \bop) - \Sigma(0, \bop)
=
(Z_\bop - 1)\ i\om + \cO(\om \ln^a \abs{\om})$
as $\om \to 0$,
with $Z_\bop$ reaching a finite value as $\Om \to 0$, see the mid plot in Fig.~\ref{fig:Z-flow}.

\begin{figure}
  \centering
   \includegraphics{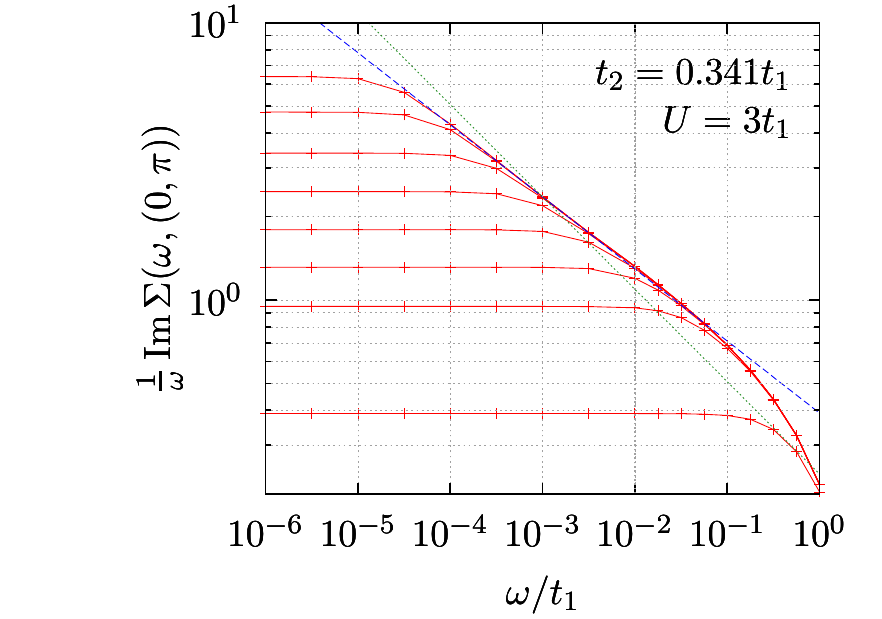}
  \caption{Small-frequency asymptotics of the self-energy at the Van Hove points.
    The discrete data shows $\rIm\Sigma$ at different stages of the flow,
    $\Om / t_1 = 10^{-5},\ {3\cdot 10^{-5}},\ 10^{-4},\ {3\cdot 10^{-4}},\ 10^{-3},\ {3\cdot 10^{-3}},\ 10^{-2}, \ 10^{-1}$
    (top-down).
    In the frequency regime $10\Om \lesssim \abs{\om} \ll t_1$
    all curves coincide.
    This allows us to extract a (non-Fermi-liquid) frequency asymptotics
    $\rIm\Sigma(\om, (0, \pi)) \simeq \textnormal{const}\ \abs{\om}^\alpha\sgn\om$ with
    $\alpha = 0.74$ (dashed line).
    For comparison, the dotted line shows the asymptotics with $\alpha = 2/3$,
    which is not met by the present data.}
  \label{fig:ImSigma_asymptotics}
\end{figure}

On the other hand, $\rIm\Sigma(\om, (0, \pi))$
gets substantial contributions from the flow at all scales $\Om$:
it grows strongly in the frequency region $\abs{\om} \lesssim \Om$
but changes little for larger $\abs{\om}/\Om$.
As a consequence,
the curves $\om \mapsto \rIm\Sigma(\om, (0, \pi))$ at different scales $\Om$ coincide
for frequencies $\abs{\om} \gtrsim 10\Om$,
see Fig.~\ref{fig:ImSigma_asymptotics},
hence provide access to the frequency dependence of the limit $\Om \to 0$.
In particular, 
the RG calculation gives the frequency asymptotics
\begin{equation} \label{eq:ImSigma_asymptotics} 
   \rIm\Sigma(\om, (0, \pi))\
\simeq
   \textnormal{const}\ \abs{\om}^\alpha\sgn\om
,
\end{equation}
with exponent $\alpha \approx 0.74$ for frequencies $\abs\om \lesssim 10^{-2}t_1$.
This value of the exponent is consistently extracted at all scales $\Om \le 10^{-4}t_1$.
The contribution \eqref{eq:ImSigma_asymptotics} dominates the $i\om$ term
in the propagator at small frequencies and leads to non-Fermi-liquid behavior.

Fermion systems at criticality have been extensively investigated in the literature.
Within the framework of a $g$-ology model,
the implications of the Van Hove scenario for a
non-Fermi-liquid frequency dependence of the two-point function
were explored.\cite{Dzyaloshinskii1996}
In the case of a regular FS with constant non-vanishing Fermi velocity
and interaction via a single bosonic channel,
the Eliashberg resummation produces the non-FL frequency asymptotics \eqref{eq:ImSigma_asymptotics}
with exponent $\alpha = 2/3$.
However, for fermionic spin SU(2) interactions the Eliashberg expansion 
is unstable with respect to vertex and 
higher-loop self-energy corrections.\cite{Rech2006}
For models of the nematic phase transition where the fermions couple to
a gapless scalar or $U(1)$ gauge field, the exponents were calculated using
three-loop field theoretic RG\cite{MelitskiSachdev2010,MrossEtal2010} 
and functional RG methods\cite{DrukierEtal2012}, all of which gave 
$\alpha < 2/3$. 
In the present situation of fermions with an aspherical FS including Van Hove points
and with competing interaction channels,
the functional form \eqref{eq:ImSigma_asymptotics} remains,
but we obtain a different exponent.

\begin{figure}
  \includegraphics{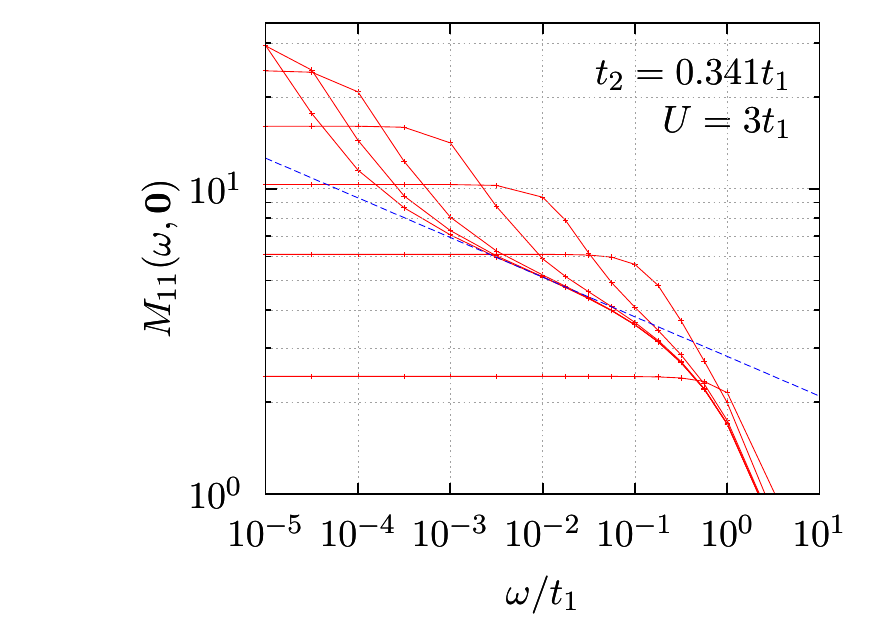}
  \caption{ 
    RG flow of the ferromagnetic exchange function  $\om \mapsto M_{11}(\om, \bzero)$
    at different stages of the flow,
    $\Om/t_1 = 10^{-5},\ 3\cdot 10^{-5},\ 10^{-4},\ 10^{-3},\ 10^{-2},\ 10^{-1},\ 10^0$
    %Right: Om = 0.00001, 0.0000291, 0.0000988, 0.0003104, 0.0009756, 0.0030660, 0.0096357, 0.0302828, 0.1027221, 0.2991048, 1.014592,
    (top-down viewed from $\om/t_1 = 10^{-5}$).
    The dashed line
    $ \sim \abs{\om}^{-\alpha_2}$
    represents the 
    estimate for the asymptotic frequency behavior of $M_{11}(\om, \bzero)$
    with the exponent $\alpha_2$ in the region $0.13 \dotsc 0.2$,
    the line uses $\alpha_2 = 0.13$.
  }
  \label{fig:FM_asymptotics}
\end{figure}

In the same way, the RG flow provides information
about the asymptotic frequency behavior in the various interaction channels.
The identification of asymptotic regimes based on data
for scales $\Om/t_1 \ge 10^{-5}$
is less obvious in the case of the four-point function
due to non-monotonic dependences on the scale parameter $\Om$ in the $\Om$ scheme:
The exchange function
    $M_{11}(\om, \bzero)$ %is a non-monotonic function of the scale parameter $\Om$:
    increases during the flow for frequencies $\abs{\om} \lesssim \Om$,
    it decreases for $\abs{\om} \gtrsim \Om$, see Fig.~\ref{fig:FM_asymptotics}.
    Assuming that the asymptotic frequency regime is already reached,
    we extract
    $M_{11}(\om, \bzero) \simeq \abs{\om}^{-\alpha_2}$
    with an exponent $\alpha_2 \approx 0.13 \dotsc 0.2$
    (green line in Fig.~\ref{fig:FM_asymptotics}).    
   Concerning the $d$-wave pairing channel, 
   it
   is hard to identify an asymptotic frequency regime from the RG data.

Varying the interaction parameter in the range $2.5 \le U/t_1 \le 3.5$
we find that the ratio $t_2/t_1$ with expected smallest $\Om_*$,
including the possibility of $\Om_* = 0$, changes only little.
The dependence of the exponent $\alpha$ on $U$ is left to future work.

%%%%%%%%%%%%%%%%%%%%%%%%%%%%%%%%%%%%%%%%%%%%%%%%%%%%%%%%%%%%%%%%%%%%%%%%%%%%%%%%%%%%%%%%%%%%%%%%%%%%%%%%%%%%%
\subsection{Comparison to flow with momentum- and frequency-dependent self-energy}

\begin{figure}
  \hspace{-2cm}
  \includegraphics{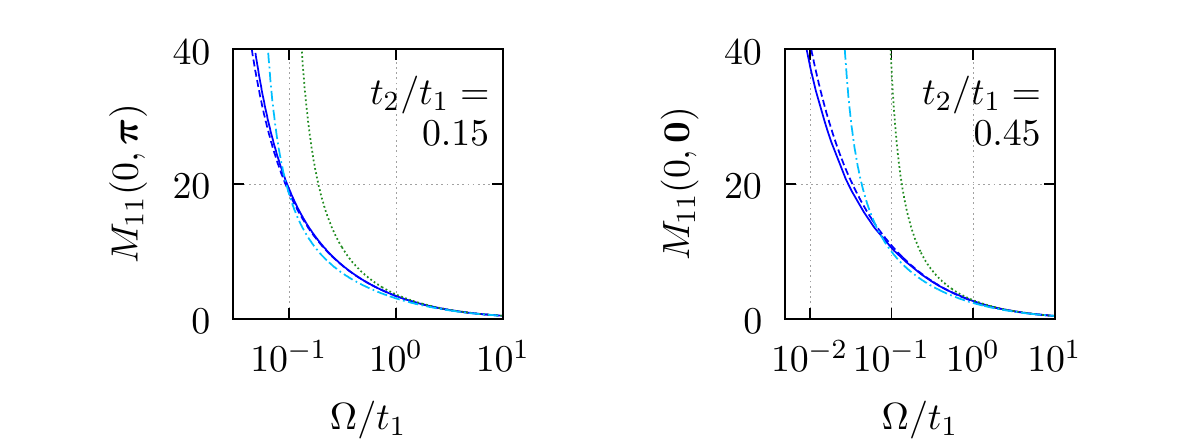}
  \caption{Flow of the dominant instability in the AFM (left) and FM regime (right)
    in different calculation schemes.  
    We compare the flow including $\rIm\Sigma(\om, \bop)$ (solid and dashed lines) 
    to several setups disregarding self-energy effects:
    frequency-independent vertex functions (dash-dot) and frequency-dependent vertex functions (dots).
    The solid line corresponds to the full flow with feed-in of 
    the momentum- and frequency-dependent $\rIm\Sigma(\om, \bop)$,
    the dashed one to a feed-in of $\rIm\Sigma(\om, (0, \pi))$ only.
    At $t_2/t_1 = 0.45$, $Z_{(0, 0)} > Z_{(0, \pi)}$ for most of the flow,
    which leads to the atypical situation that the momentum-independent feed-back of $\rIm\Sigma(\om, (0, \pi))$
    slightly overestimates the true flow, instead of underestimating it.}\label{fig:exampleFlows_wFreq-wSig}
\end{figure}

Because of numerical complexity
the flow with momentum-dependent $\rIm\Sigma$  was 
calculated only for parameter values $0 \le t_2/t_1 \le 0.15$ and $0.45 \le t_2/t_1 \le 0.5$.
Here, the approximation of feed-in of momentum-independent $\rIm\Sigma(\om, (0, \pi))$
to the flow turns out to produce only a small error:
Figure~\ref{fig:exampleFlows_wFreq-wSig} 
compares the flow of the dominant coupling for
$t_2/t_1 = 0.15$ (AFM regime) and $0.45$ (FM regime),
only a small correction is found
when accounting for the momentum dependence of $\rIm\Sigma$.

For the remaining parameter range $0.15 < t_2/t_1 < 0.45$, where flows
go down to small scales, a larger error is expected.
Although the full flow could not be obtained in this region,
full scale derivatives for specific couplings
can still be calculated here.
This serves as an error estimate to the approximated flow with 
momentum-independent but frequency-dependent feed-back of $\rIm\Sigma(\om, \boq)$,
$\boq = (0, \pi)$.
Using the data obtained in this way, we compare full and approximated
scale derivatives for the most dominant couplings at the stopping scale.
Since $(\partial_\om \rIm\Sigma)(\om = 0, \bop)$ typically grows strongest
for $\bop = (0, \pi)$ the approximated flow is usually more suppressed than the true flow. 
Since 
$\abs{ \rIm\Sigma(\om, \bop) - \rIm\Sigma(\om, (0, \pi) ) }$
typically increases 
during the flow,
the comparison of scale derivatives at the end of the flow yields 
the largest error in the scale derivatives during the whole flow.

For parameter values $t_2/t_1 = 0.15, 0.3, 0.33$ and $0.42$ we find that
scale derivatives of the leading and next-to-leading couplings 
of the interaction vertex get, 
at the stopping scale and by feed-in of momentum-dependent $\rIm\Sigma$,
a relative correction of $1\%..15\%$. 

\begin{figure}
 \includegraphics{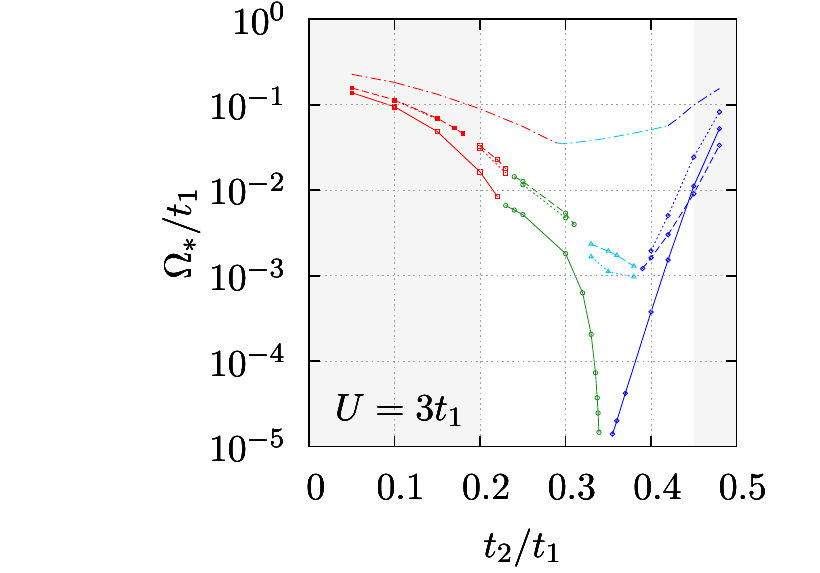}
   \caption{Approximation of momentum-independent feed-back of 
$\rIm\Sigma(\om, \boq)$ to the flow. The results for several choices of $\boq$
 (solid: $\boq = (0, \pi)$, dashed: $\boq = (0, 0)$, dots: $\boq = (\pi, \pi)$)
share three common characteristics compared to neglect of self-energy (uppermost curve):
The stopping scale gets (for intermediate $t_2/t_1$ substantially) suppressed;
a $d$-SC dominated parameter region is recovered;
the scattering process with non-zero frequency exchange is weakened.\\
The flow with full $\rIm\Sigma$ feed-in is expected to be close to the 
approximate one with $\boq = (0, \pi)$, but with a higher stopping scale.
On the other hand, from the discussion in the text the true stopping scale is supposed
to be well below the ones obtained from choices $\boq = (0, 0)$ or $(\pi, \pi)$.
Color conventions as in Fig.~\ref{fig:Omcrit_ImSigma-flow}.}
   \label{fig:Omcrit_differentq}
\end{figure}

Alternative choices $\boq = (0, 0)$ or $\boq = (\pi, \pi)$ underestimate
$\rIm\Sigma$ in the saddle point region and lead to larger stopping scales,
see Fig.~\ref{fig:Omcrit_differentq}. 
With such choices of $\boq$, scale derivatives are highly overestimated,
by a factor of $2..5$,
which underlines the special role of the saddle point region for systems at VHF.

From this check it can be deduced that
(a) accounting for the correct $\rIm\Sigma$ in the saddle point region
is of major importance at VHF
and (b) the true flow with full $\rIm\Sigma$ should be close to the one
with $\rIm\Sigma(\om, (0, \pi))$ feed-back.

%%%%%%%%%%%%%%%%%%%%%%%%%%%%%%%%%%%%%%%%%%%%%%%%%%%%%%%%%%%%%%%%%%%%%%%%%%%%%%%%%%%%%%%%%%%%%%%%%%%%%%%%%%%%%
\subsection{Effect of the imaginary part of exchange propagators}

Already in perturbation theory a singularity in the imaginary part of the interaction vertex
is found: 
Although the imaginary part of the particle--particle bubble vanishes at zero frequency,
it shows a singular frequency derivative.
For zero momentum transfer, at VHF and with $0 < t_2/t_1 < \undemi$,
\begin{equation}
    \partial_{p_0} \Big|_{p_0 = 0}\ \rIm \int\intd l\ C_\Om(-l) C_\Om(p_0 + l_0, \bol)
=
    - c\ \ln^2\Om + \cO(\ln\Om),
\qquad c > 0.
\end{equation}
This singularity is induced by the asymmetry of the density of states $\cN(e)$
about energy $e = 0$.
Upon ladder resummation in the SC channel, an even stronger singularity
$\sim (\Om - \Om_*)^{-2}$ builds up in this frequency derivative
when the composite pair field becomes massless.

\begin{figure}
  \includegraphics{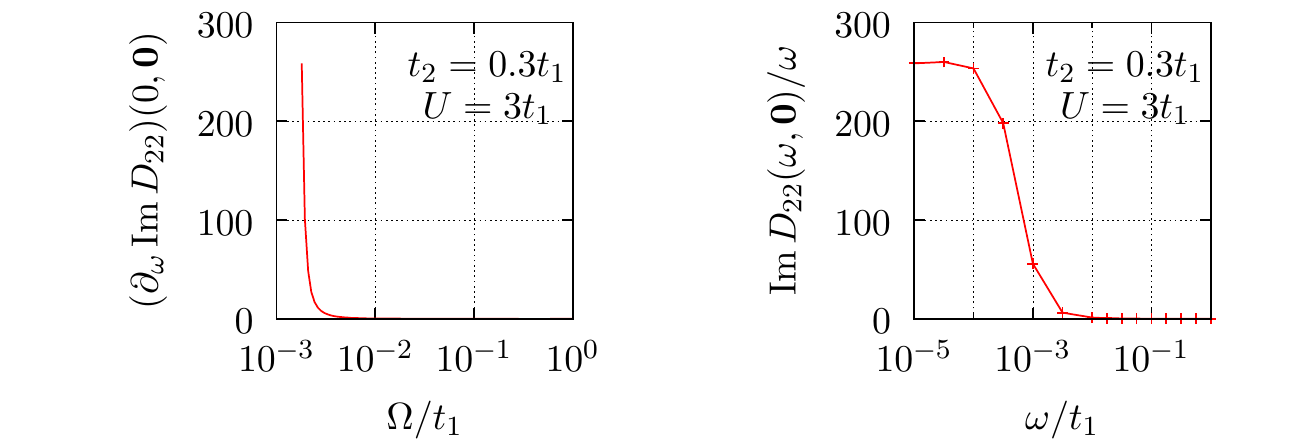}
  \caption{Imaginary part of the $D_{22}$ exchange function.
    Left: Flow of its first frequency derivative.
    Right: The function $\om \mapsto \fracun{\om} \rIm D_{22}(\om, \bzero)$ 
      at the stopping scale $\Om_* \approx 0.002 t_1$.}\label{fig:ImD22_Form}
\end{figure}

In our parametrization, only $D_{11}$ and $D_{22}$ can acquire an imaginary part, 
see the symmetry discussion below eq.~\eqref{eq:B_symmetries}.
We have already examined the influence of this singularity on the flow of the
frequency-dependent interaction vertex with neglect of self-energy\cite{HGS2011} at the data point
$t_2/t_1 = 0.3,\ U/t_1 = 3$. This influence was found to be minimal:
despite the singular frequency derivative, imaginary exchange propagators remain small during flow.
For the same data point, we study here the influence of $\rIm D_{11}$, $\rIm D_{22}$ on the
combined flow of the frequency-dependent interaction vertex and $\rIm\Sigma(\om, (0, \pi))$.
In this parameter region of dominant $d$-SC, AFM correlations are found to generate a strong
$(\partial_\om \rIm D_{22})(\om = 0, \bzero)$, 
which becomes much greater than the frequency derivative of $\rIm D_{11}$, 
see Fig.~\ref{fig:ImD22_Form}.
Nevertheless the functions $\rIm D_{mm}(l)$ remain small during flow, as found before. 

\begin{figure}
  \includegraphics{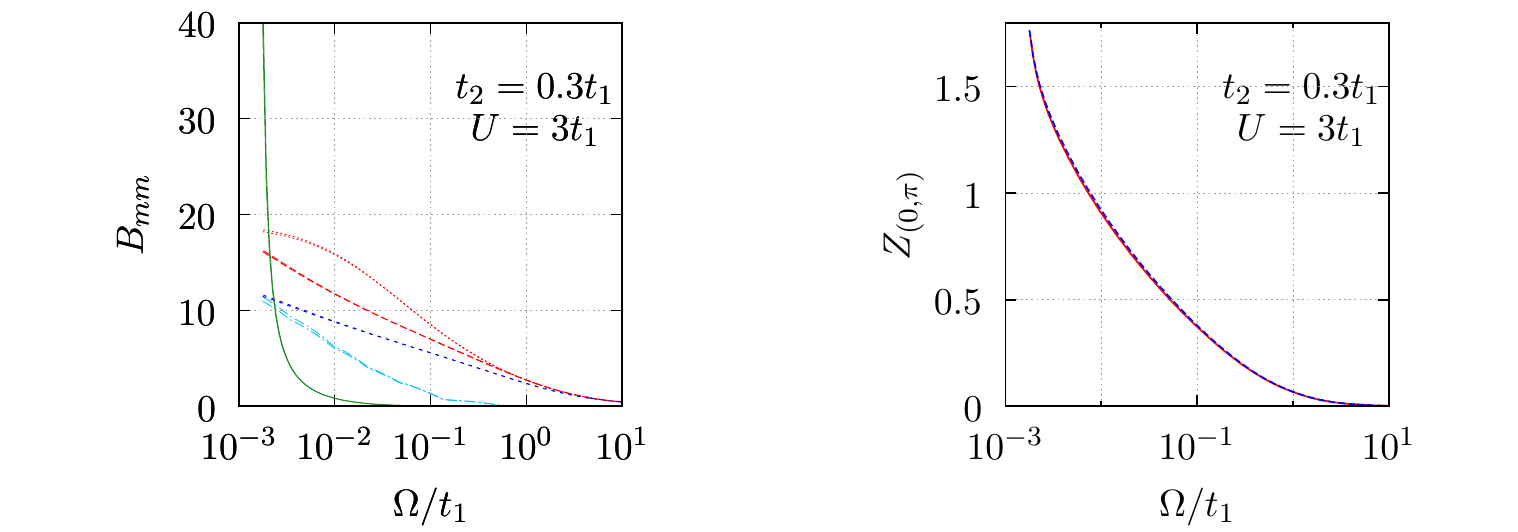}
   \caption{Comparison of flows taking into account $\rIm D_{11}$ and $\rIm D_{22}$
     or neglecting them. 
     Left: Flow of the leading couplings, 
     line conventions as in Fig.~\ref{fig:Omcrit_ImSigma-flow}.
     Right: Flow of $Z_{(0, \pi)}$.
     Both flows are almost indistinguishable.}
  \label{fig:Influence_ImDmm}
\end{figure}

Accordingly, these imaginary vertex contributions 
practically do not change
the flow of $\rIm\Sigma$
and the most singular couplings, see Fig.~\ref{fig:Influence_ImDmm}.
We leave the calculation of the flow with $\rIm D_{mm}$ in other parameter regions
for future work but do not expect a strong effect there either.

%%%%%%%%%%%%%%%%%%%%%%%%%%%%%%%%%%%%%%%%%%%%%%%%%%%%%%%%%%%%%%%%%%%%%%%%%%%%%%%%%%%%%
\section{Functional form of frequency dependences in the symmetric phase}
\label{sec:transfer-freq-dep}

The functional form of the frequency dependence of exchange propagators
and of the self-energy is of particular interest
in view of going beyond the numerically expensive frequency discretization used so far.
Based on the discrete RG data we now propose and discuss several parametrizations.

RPA calculations motivate a transfer frequency dependence in form of a Lorentz 
curve,\cite{HGS2011}
%which can be chosen 
such that the small-frequency region is captured accurately.
However, we find that in the intermediate to large frequency region 
$\abs{\om} \gtrsim \Om$ 
the true frequency behavior 
deviates from simple Lorentzians,
especially at low RG scales,
and that the intermediate frequency region $\abs{\om} \approx \Om$ 
can be of considerable importance for the calculation of a quantitatively
correct flow.

Generalizations of single Lorentz curves
considerably extend
the regime of validity of the functional parametrization for 
transfer frequency dependences $\om \mapsto B_{mm}(\om, \bop)$
and for the frequency dependence of the self-energy in the form
$\om \mapsto \om^{-1}\rIm(\om, \bop)$:
At not too low RG scales $\Om$, a sum of two Lorentzians
\begin{equation}\label{eq:freqfit_2L}
  f(\om)
=
  \frac{a_1}{1 + b_1^2\, \om^2} + \frac{a_2}{1 + b_2^2\, \om^2}
\end{equation}%
captures frequency dependences quite well.\cite{HGS2011}
These two curves have the interpretation of a small- and a large-frequency process.
Furthermore, in the frequency region $\abs\om \le 10\Om$
the fit to two Lorentzians
is very accurate at all scales $\Om / t_1 \ge 10^{-5}$;
this frequency region is most important in the calculation of the RG flow
of the leading and sub-leading couplings.

Whereas the exchange functions $M_{11}$, $D_{11}$ and $D_{22}$
in the magnetic and pairing channel remain positive for all
frequency-momenta
such that the corresponding fermionic interaction can be decoupled by a %an appropriate
Hubbard-Stratonovich transformation,
the RG flow detects a sign change in the scattering channel
around zero transfer momentum
with a maximum of $K_{11}(\om, \bzero)$ at $\om = 0$
and a pronounced minimum $K_{11}(\om_{\textnormal{s}}, \bzero) < 0$ at $\om_{\textnormal{s}} > 0$,
see Ref.~\onlinecite{HGS2011}.
In the parametrization with Lorentzians,
at least two Lorentz curves are needed to account for this behavior.

\begin{figure}
  \centering
  \includegraphics{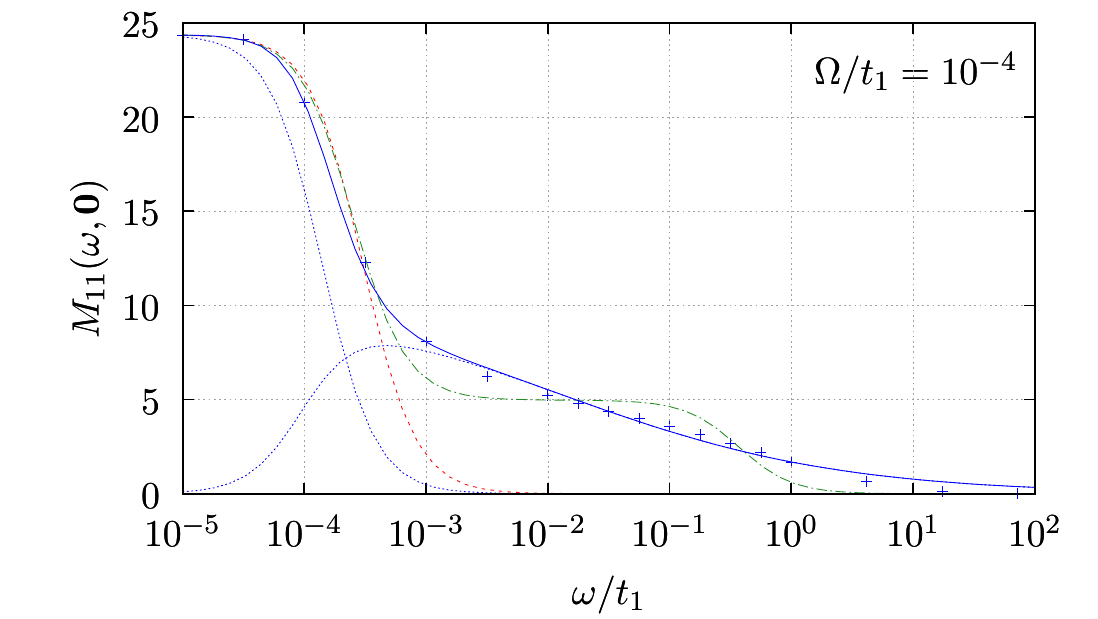}
  \caption{Overview about the various functional parametrisations
    for the transfer frequency dependence of exchange propagators at low RG scales.
    On the example of the ferromagnetic exchange $\om \mapsto M_{11}(\om, \bzero)$
    several ansatzes for the discrete RG data ($+$ symbols) are compared:
    A single Lorentzian determined from the small-frequency behavior (red dashed line)
    is appropriate for small frequencies $\abs\om \ll \Om$.
    Likewise, the sum of two Lorentzians (green line in dashes and dots) 
    is a compromise to the different behavior in different frequency regimes
    and not an accurate parametrization for a large frequency range.
    The functional form \eqref{eq:freqfit_La2} leaves the frequency exponent
    in the region of intermediate frequencies free and describes the RG data quite well (blue solid line).
    It decomposes into the two blue dotted lines.
    The parameter values are $t_2 / t_1 = 0.341$, $U/t_1 = 3$.
  } 
  \label{fig:freqfit_La2}
\end{figure}

\bigskip
The parameter range $t_2/t_1 \approx 0.34$
exhibits very low stopping scales,
and the ferromagnetic exchange function 
as well as the self-energy at the Van Hove points
develop a characteristic 
%non-Fermi-liquid 
frequency asymptotics at small frequencies,
see Sec.~\ref{non-FL-Sigma}.
In this case, different frequency regimes occur:
In the small-frequency region $\abs{\om} \ll \Om$,
the flow in the $\Om$ scheme detects at all scales a quadratic dependence
as \eg in Lorentzians;
this result possibly depends on the regularization method.
On the other hand,
a pronounced intermediate frequency region with a characteristic decay
$\abs{\om}^{-\gamma}$ forms during flow,
consider \eg the region $10^{-3} \le \om/t_1 \le 10^{-1}$ in Fig.~\ref{fig:freqfit_La2}.
In the $\Om$ scheme, this decay behavior 
no more changes at decreasing $\Om$ for frequencies $\abs{\om} \gtrsim 10 \Om$,
and we therefore expect it to be independent of the regularization
and to provide the small-frequency asymptotics as the regularization is fully removed.

In this context, a suitable modification of \eqref{eq:freqfit_2L} is an ansatz
where the frequency exponent $\gamma > 0$ describing the intermediate
frequency region can be varied,
\begin{equation}\label{eq:freqfit_La2}
  f(\om)
=
  \frac{a_1}{1 + b_1^2\, \om^2} + \frac{a_2}{1 + b_2^2\, \abs\om^\gamma}\ \chi_\Om(\om)
.
\end{equation}
Fig.~\ref{fig:freqfit_La2} shows the relatively high quality that is achieved
by this ansatz,
as compared to the parametrizations with one or two Lorentzians.
All fits are determined with a least-squares method and a ten times larger
weighting factor for frequencies $\abs{\om} \le \Om$.
The frequency data obtained in the RG flow has $0 < \gamma < 1$
at low RG scales $\Om$.
A factor $\chi_\Om(\om)$ is included in the second term of \eqref{eq:freqfit_La2}
in order to satisfy $f'(\om = 0) = 0$;
this condition is imposed on the exchange propagators and on the function
$\om^{-1}\rIm\Sigma(\om, \bop)$
by the $\Om$ regularization at scales $\Om > 0$.
For simplicity,
the $\Om$ regulator $\chi_\Om(\om) = \om^2 / (\om^2 + \Om^2)$
is used in the ansatz.
Apart from the scattering exchange $K_{11}(p)$, which has a sign change
along the frequency axis,
we find that exchange propagators at all scales $\Om/t_1 \ge 10^{-5}$
are well captured by \eqref{eq:freqfit_La2}.

\begin{figure}
  \hspace*{.5cm} 
  \includegraphics{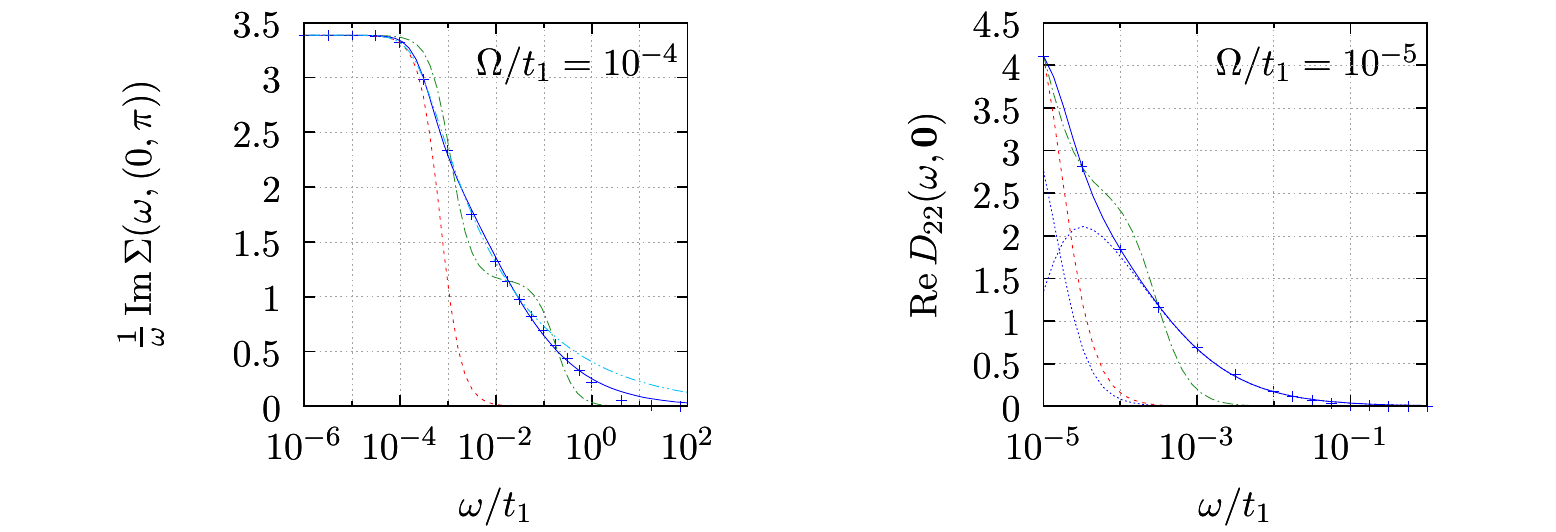}
    \caption{ Effective functional forms from the discretized frequency dependences 
      obtained with the RG flow for
      $\om \mapsto \om^{-1}\rIm \Sigma(\om, (0, \pi))$
      and
      $\om \mapsto \rRe D_{22}(\om, \bzero)$.
      The line conventions and further details are given in Fig.~\ref{fig:freqfit_La2}.
      The fit for $\om^{-1}\rIm\Sigma$ uses the additional freedom 
      provided by the ansatz \eqref{eq:freqfit_La3} instead of \eqref{eq:freqfit_La2};
      the cyan dash-dot-dot line shows the fit to the three-parameter ansatz \eqref{eq:freqfit_La5}
      and captures well the discrete data except for the large-frequency asymptotics.
      The corresponding fit parameters are given in Tables \ref{tab:freqfit1}
      and \ref{tab:freqfit2}.
      Data at frequency $\om = 0$ (not shown in the logarithmic plots)
      is very close to the values at the lowest frequencies given in the graphs,
      in agreement with the Lorentzian form at very small frequencies.
      The parameter values are $t_2 / t_1 = 0.341$, $U/t_1 = 3$.
    } 
  \label{fig:freqfit_overview}
\end{figure}

The frequency-dependent self-energy in the form $\om \mapsto \om^{-1}\rIm\Sigma(\om, \bop)$
is described well at all scales $\Om/t_1 \ge 10^{-5}$
by the generalization of \eqref{eq:freqfit_La2} to
\begin{equation}\label{eq:freqfit_La3}
  f(\om)
=
  \frac{a_1}{1 + b_1^2\, \om^2} + \frac{a_2}{1 + b_2^2\, \abs\om^\gamma}\ \frac{\om^2}{\om^2 + c^2}
,
\end{equation}
where $\chi_c(\om)$ replaces $\chi_\Om(\om)$
and $c > 0$ is a further variational parameter.
The ansatz
\begin{equation}\label{eq:freqfit_La5}
  f(\om)
=
  \frac{a}{(1 + b^2\, \om^2)^{\gamma/2}}
\end{equation}
is designed to capture the quadratic dependence at small frequencies
as well as the characteristic decay at intermediate frequencies
with just three parameters.
Whereas this ansatz does not reproduce the various transfer frequency dependencies,
we find that \eqref{eq:freqfit_La5}
indeed describes $\om^{-1} \rIm\Sigma$ very well
in the small to intermediate frequency region
and hence is particularly suitable to account for
self-energy effects when calculating the flow of the leading exchange couplings.
Figures \ref{fig:freqfit_La2} and \ref{fig:freqfit_overview} 
show the quality of these ansatzes,
Tables~\ref{tab:freqfit1} and \ref{tab:freqfit2} provide the corresponding fit parameters.

%% commands for "scientific" exponential notation etc in typewriter font (saves space)
%\newcommand{\myE}[2]{\textnormal{\texttt {#1}E{#2}}\hspace{1em}}
\newcommand{\myE}[2]{{\texttt {#1}\cdot 10^{#2}}\hspace{.8em}}
\newcommand{\myN}[1]{\textnormal{\texttt {#1}}}

\begin{table}
  % hspace: in equation environment below
  {\small
  \begin{equation*}
%  \hspace*{-2cm}
  \renewcommand{\arraystretch}{1.15}
  \begin{array}{|l | l l l l | l l l l | }
   \hline
   & \multicolumn{4}{|c|}{M_{11}(\om, \bzero)} & \multicolumn{4}{|c|}{\rRe D_{22}(\om, \bzero)} \\
  \hline
   \Om & 10^{-2} & 10^{-3} & 10^{-4} & 10^{-5} &     10^{-2} & 10^{-3} & 10^{-4} & 10^{-5}  \\        
  \hline
   a_1 & \myE{1.0}{+1} & \myE{1.6}{+1} & \myE{2.4}{+1} & \myE{3.2}{+1} &         \myE{8.6}{-2} & \myE{4.5}{-1} & \myE{1.4}{+0} & \myE{4.1}{+0} \\     
   b_1 & \myE{7.4}{+1} & \myE{6.9}{+2} & \myE{7.1}{+3} & \myE{7.0}{+4} &         \myE{2.5}{+1} & \myE{1.1}{+3} & \myE{1.1}{+4} & \myE{6.9}{+4} \\     
   a_2 & \myE{5.5}{+0} & \myE{6.9}{+0} & \myE{1.1}{+1} & \myE{3.5}{+2} &        \myE{-2.3}{-2} & \myE{4.0}{-1} & \myE{1.4}{+0} & \myE{3.1}{+0} \\     
   b_2 & \myE{1.5}{+0} & \myE{1.8}{+0} & \myE{2.3}{+0} & \myE{1.3}{+1} &         \myE{9.3}{+1} & \myE{1.7}{+1} & \myE{1.7}{+1} & \myE{2.0}{+1} \\     
   \gamma & \myN{0.84} & \myN{0.58}& \myN{0.38}& \myN{0.19} &                                  \myN{2.8} & \myN{1.1} & \myN{0.78} & \myN{0.68} \\   
   \hline
  \end{array}
\end{equation*}
  } % end of small
  \caption{Fitting parameters for the discretized frequency dependence 
    of the ferromagnetic exchange $M_{11}(\om, \bzero)$
    and the $d$-SC exchange $D_{22}(\om, \bzero)$,
    as obtained in the RG flow.
    At the considered parameter point $t_2 / t_1 = 0.341$, $U/t_1 = 3$
    these exchange channels are the most important ones.
    The fits to the ansatz \eqref{eq:freqfit_La2} 
    are provided at four different RG scales $\Om/t_1 = 10^{-2} \dotsc 10^{-5}$.
    Figures \ref{fig:freqfit_La2} and \ref{fig:freqfit_overview} 
    show the quality of the fits.
    All the data in the table uses units where $t_1 = 1$.
  }
  \label{tab:freqfit1}
\end{table}

\begin{table}
  % hspace: in equation environment below
  {\small
  \begin{equation*}
  \hspace*{-2cm}
  \renewcommand{\arraystretch}{1.15}
  \begin{array}{|l | l l l l l l |}
   \hline
   & \multicolumn{6}{|c|}{\om^{-1}\rIm\Sigma(\om, (0, \pi))}\\
  \hline
   \Om & 10^{-2} & 3\cdot 10^{-3} & 10^{-3} & 3\cdot 10^{-4} & 10^{-4} & 10^{-5} \\
  \hline
   a & \myE{9.4}{-1} & \myE{1.3}{+0} & \myE{1.8}{+0} & \myE{2.5}{+0} & \myE{3.4}{+0} & \myE{6.4}{+0} \\
   b & \myE{2.3}{+1} & \myE{8.8}{+1} & \myE{3.1}{+2} & \myE{1.2}{+3} & \myE{4.2}{+3} & \myE{4.3}{+4} \\
   \gamma & \myN{0.38} & \myN{0.30}      & \myN{0.28}      & \myN{0.25}      & \myN{0.25}      & \myN{0.26}  \\
  \hline
  \end{array}
\end{equation*}
  } % end of small
  \caption{Fitting parameters for the discretized self-energy frequency dependence 
    $\om \mapsto \om^{-1} \rIm\Sigma(\om, (0, \pi))$,
    as obtained in the RG flow.
    The table gives the parameters for the ansatz \eqref{eq:freqfit_La5}
    at several RG scales,
    the quality of this fit is shown in Fig.~\ref{fig:freqfit_overview}.
    All the data in the table uses units where $t_1 = 1$.
  }
  \label{tab:freqfit2}
\end{table}

Notice that the fit parameters $\gamma$ are related 
to the exponents identified in Sec.~\ref{non-FL-Sigma}:
the parameter $\gamma$ in the ferromagnetic exchange propagator corresponds
to the exponent $\alpha_2$ given previously;
the self-energy parameter $\gamma$ 
corresponds to $1 - \alpha$,
 since the fit uses an additional factor $\om^{-1}$.
The fit parameters $\gamma$ in Tab.~\ref{tab:freqfit1}
are very close to
the exponents determined in Sec.~\ref{non-FL-Sigma},
however do not always coincide exactly. 
This is because $\gamma$ here is determined as the best compromise to describe
the behavior for all frequencies,
whereas the true exponent characteristic to the asymptotic behavior at small frequencies
must be extracted exclusively from the frequency region
where the asymptotics shows.

\bigskip
It is interesting to note that
the flow of the leading couplings 
(including their vicinity in frequency-momentum space)
and the flow of the self-energy at small frequencies
is in many cases sensitive only 
to proper parametrization of a small frequency region $\abs\om \lesssim \Om$,
especially at not too low RG scales.
The behavior of exchange propagators and the self-energy
in this small frequency region
can be described well with a sum of two Lorentzians
or, at not too low RG scales, even by a single Lorentz curve.
The more general ansatzes \eqref{eq:freqfit_La2} to \eqref{eq:freqfit_La5}
capture well the behavior in a much larger frequency region
at all scales $\Om/t_1 \ge 10^{-5}$
and allow to describe the forming of the specific asymptotic 
frequency dependence as $\Om \to 0$.

The generalization of the functional form of the frequency dependence
from a single Lorentz curve to (real-valued) functions
\eqref{eq:freqfit_2L} -- \eqref{eq:freqfit_La5}
raises the question of how to extract the free parameters 
in the ansatz from the flow equation:
while in these ansatzes 
two parameters can in principle be fixed by the Taylor
expansion at zero frequency, the remaining parameters
account for the behavior in the intermediate-frequency region.
In the present work they are determined from a least-squares fit of discrete data.
A more general way of extracting a flow equation for them needs yet to be established.
With an appropriate method for setting
up the flow equations for these parameters, 
the numerical complexity of the evaluation of the flow equations \eqref{eq:Sigma_v_flow} will
be substantially lowered and further insight in the structure of their solution
can be gained.

%%%%%%%%%%%%%%%%%%%%%%%%%%%%%%%%%%%%%%%%%%%%%%%%%%%%%%%%%%%%%%%%%%%%%%%%%%%%%%%%%%%%%%%%%%%%%%%%%%%%%%%%%%%%%%%%%%%%%%%
\section{Conclusions}

In this work we have  solved the level-2 truncated RG equations for the 
two-dimensional repulsive Hubbard model,  
including the feed-back of the self-energy to the vertex flow equations 
via the full propagators in the flow equations. 
We have computed the stationary self-energy and thus determined 
the band function of the interacting system, and, in a second step, 
calculated the imaginary part of the self-energy. 
To our knowledge this is the first time in this two-dimensional model 
that the frequency dependence of vertex functions and self-energy have 
been taken into account without simplifying assumptions on their
functional form, while also keeping the momentum dependence in a 
well-tested approximation.  Although our method is applicable in general, 
i.e.\ without restriction to
particular parameter regimes, we have focused on the situation where
the interacting system is at Van Hove filling, because of the intrinsic theoretical
interest of this situation and because of the indications from ARPES measurements
that the Fermi surfaces of cuprates are indeed close to Van Hove points. 

The stationary self-energy determines the band dispersion of the electrons in
the interacting system. 
In the RG setup, 
it becomes a scale-dependent quantity, hence the Fermi surface also flows under the RG.
This is a power counting relevant effect at weak interactions, hence a priori important. 
Capturing this flow is technically nontrivial in momentum cutoff schemes, 
but much easier with the frequency regulator that we use here.
We have Fourier expanded the self-energy into  a sum of correction terms
to free electron hopping and fixed the particle density during the flow 
by adjusting the self-energy zero mode.  By proper choice of the chemical potential,
we tuned the density to interacting Van Hove filling.

The corrections to the hopping amplitudes were calculated by orthogonal projection
of the self-energy to standard hopping functions.
This method was found to be consistent with an alternative one
that extracts coefficients from a Taylor expansion of the self-energy around the 
Van Hove points, supplemented by conditions obtained from the self-energy 
at other points in momentum space. We also found that  the attempt
to use only local information at the Van Hove points leads to 
unstable behavior, i.e.\ including higher terms in the Taylor expansion around
these points leads to big changes in the coefficients determined from the lower orders. 
This problem disappears when an overemphasis on the Van Hove points is avoided
by taking into account other momentum space points.

For the  parameter values considered here and with our fixed-density constraint, 
we found only a small change of the Fermi surface in the flow, resulting only in minor modifications
of the flow of the interaction vertex.

Our results on the imaginary part of the self-energy allow us to study 
several aspects of the frequency-dependent two-point function,
most importantly the decrease of quasi-particle
weights in the interacting system.
We find that away from frequencies below or around the RG scale $\Omega$
the linear frequency parametrization of the self-energy,
as obtained by Taylor expansion around frequency zero,
is inappropriate and leads to an artificial suppression of the flow.
We therefore discretize the self-energy behavior in frequency space,
putting special emphasis on the small-frequency region.
In momentum space, the neighborhood of the Van Hove points is particularly important:
perturbation theory finds a singularity in the first frequency derivative of the self-energy,
and the RG calculation gives an enhancement of that singularity.
This momentum space region also substantially drives the flow of the interaction vertex,
and hence properly taking into account the self-energy here is crucial.

The above-mentioned problems with the accuracy of Taylor expansions are not surprising, 
especially for self-energies with a singular frequency behavior: Taylor expansion always
works at very small scales, i.e.\ when $|\omega/\Omega|$ is very small; 
this is in the nature of a regularized theory, where singularities are smoothed out by
the regulator. It is, however, another matter to extend this to larger frequencies
$|\omega| \ge \Omega$, and our results show that the above-mentioned Taylor expansion 
around $\omega =0$ does not represent the function in the frequency range
above $\Omega$ correctly. 

We have used our numerical results to fit specific simple functional forms
to the self-energy and vertex functions. 
Our ansatzes provide a considerably larger regime of validity than simple Lorentzians
coming from Taylor expansion, 
and will be of interest in future studies of the model, as they allow to avoid
a fully numerical study while retaining the accuracy of the frequency dependence. 
The coefficients can again be determined with much less numerical effort 
(but -- again -- not only by a local Taylor expansion that would overemphasize 
the vicinity of the singularity).
The behavior at very large frequencies is given by a convergent expansion in 
inverse powers of $\omega$, 
and may have coefficients differing from those we get in the intermediate frequency regime. 
This leads, however, only to small changes.

Concerning the application to the Hubbard model, 
we find that at the Van Hove filling,  as a consequence of decreasing quasi-particle weights,
the flow of the frequency-dependent interaction vertex gets slowed down.
This effect is most drastic
in the parameter region of competing $d$-wave pairing and ferromagnetic instabilities, 
where the stopping scale of the flow drops by several orders of magnitude,
 confirming earlier suggestions of a quantum critical point.
At this point we calculate the non-Fermi-liquid frequency dependence
of the symmetric self-energy. It is a fractional power law, as found previously in models
without Van Hove singularities, but with a different exponent.

In comparison to the results from flows with a frequency-dependent vertex function
where the self-energy is neglected,\cite{HGS2011} the stopping scale for the flow
of the frequency-dependent interaction vertex with self-energy feedback 
drops below that obtained in the static approximation.\cite{HonerkampSalmhofer_Tflowletter,HusemannSalmhofer2009}
%where the frequency dependence is neglected also in the vertex function. 
Similarly, a region of dominant $d$-wave pairing reappears also at $U=3t_1$, 
and scattering processes with non-zero frequency exchange get weakened, 
consistent with the fact that the $d$-wave correlations also get driven by 
contributions from the zone diagonal, while the magnetic ones depend much 
more strongly on the behavior close to the Van Hove points. 

In summary, as far as the stopping scale and dominant correlations are concerned, 
the fully frequency-dependent flow calculated here agrees well with the results from our 
flows with frequency-independent functions. This is a reassuring indication that
the projection to zero frequency, which gives the leading behavior in 
weak-coupling power counting, really works at the values of $U$ that we 
consider. Our detailed analysis reveals, however, that this is not to be understood via
a trivial Taylor-expansion argument, but that a more subtle mechanism is at work,
namely that the decrease of the quasi-particle weights works against the relative enhancement 
of the vertex at zero frequency, which in turn is caused by the decay of the vertex function in frequency.

The deviations appearing in the approximation where the 
vertex frequency dependence is kept but the self-energy is dropped
are not surprising in hindsight; vertex function and self-energy are linked by a Ward identity
and a field equation. It is more plausible that one can approximately keep these relations
either by dropping the frequency dependence of both functions, or by keeping it for 
both, than by going only half-way. At this level of generality, 
this argument does, however, not explain why the static approximation also fits
quantitatively for the above-mentioned quantities. 

A very interesting theoretical question is whether
there is a truly deconfined quantum critical point at the 
transition from superconductivity to ferromagnetism. 
Our results are consistent with this, and they imply that if there were an ordered phase, 
it would appear only at a tiny scale:
in the RG flow, the growth of different terms in the interaction competes with the suppression
by self-energy effects at the Van Hove points. At $t_2/t_1 \approx 0.34$, 
the growth tendencies of $d$-wave pairing and ferromagnetic correlations cancel one another, 
leaving the suppression of the quasi-particle weight as the dominant effect, which drives a 
drastic downturn of the stopping scale at that point and, in absence of further competition,
will suppress that scale %, and hence extend the symmetric phase, 
to zero. 
%We have found no immediately apparent competing effect
%(a $p$-wave triplet pairing is suppressed
%kinematically because the gap function has nodes at the saddle points).
One possible shielding of the QCP would be the appearance of a $d$-wave phase which 
is exclusively driven by the vicinity of the zone diagonals. 
At very weak coupling $U$, 
it could appear only below scales of the order $\exp(-\frac{1}{U^2})$.
We have also not seen it  at the values of $U$ and in the scale regime we study.
A more detailed investigation using the effective action we derived here is under way.

%%%%%%%%%%%%%%%%%%%%%%%%%%%%%%%%%%%%%%%%%%%%%%%%%%%%%%%%%%%%%%%%%%%%%%%%%%%%%%%%%%%%%%%%%%%%%%%%%%%%%%%%%%%%%%%%%%%%%%%
\section*{Acknowledgments}

We thank C. Honerkamp, C. Husemann, A. Eberlein, S. Friederich, W. Metzner, S. Uebelacker,
and A. Isidori for discussions. 
This work was supported by the DFG research unit FOR 723.

%%%%%%%%%%%%%%%%%%%%%%%%%%%%%%%%%%%%%%%%%%%%%%%%%%%%%%%%%%%%%%%%%%%%%%%%%%%%%%%%%%%%%%%%%%%%%%%%%%%%%%%%%%%%%%%%%%%%%%%
%% bibliography
%\newpage
\bibliographystyle{apsrev}
\bibliography{bib}

\begin{thebibliography}{52}
\expandafter\ifx\csname natexlab\endcsname\relax\def\natexlab#1{#1}\fi
\expandafter\ifx\csname bibnamefont\endcsname\relax
  \def\bibnamefont#1{#1}\fi
\expandafter\ifx\csname bibfnamefont\endcsname\relax
  \def\bibfnamefont#1{#1}\fi
\expandafter\ifx\csname citenamefont\endcsname\relax
  \def\citenamefont#1{#1}\fi
\expandafter\ifx\csname url\endcsname\relax
  \def\url#1{\texttt{#1}}\fi
\expandafter\ifx\csname urlprefix\endcsname\relax\def\urlprefix{URL }\fi
\providecommand{\bibinfo}[2]{#2}
\providecommand{\eprint}[2][]{\url{#2}}

\bibitem[{\citenamefont{Emery}(1987)}]{Emery1987}
\bibinfo{author}{\bibfnamefont{V.~J.} \bibnamefont{Emery}},
  \bibinfo{journal}{Phys. Rev. Lett.} \textbf{\bibinfo{volume}{58}},
  \bibinfo{pages}{2794} (\bibinfo{year}{1987}).

\bibitem[{\citenamefont{Markiewicz}(1997)}]{Markiewicz}
\bibinfo{author}{\bibfnamefont{R.}~\bibnamefont{Markiewicz}},
  \bibinfo{journal}{J. Phys. Chem. Solids} \textbf{\bibinfo{volume}{58}},
  \bibinfo{pages}{1179} (\bibinfo{year}{1997}).

\bibitem[{\citenamefont{Damascelli et~al.}(2003)\citenamefont{Damascelli,
  Hussain, and Shen}}]{Damascelli_ARPES}
\bibinfo{author}{\bibfnamefont{A.}~\bibnamefont{Damascelli}},
  \bibinfo{author}{\bibfnamefont{Z.}~\bibnamefont{Hussain}}, \bibnamefont{and}
  \bibinfo{author}{\bibfnamefont{Z.-X.} \bibnamefont{Shen}},
  \bibinfo{journal}{Rev. Mod. Phys.} \textbf{\bibinfo{volume}{75}},
  \bibinfo{pages}{473} (\bibinfo{year}{2003}).

\bibitem[{\citenamefont{{Kordyuk} and {Borisenko}}(2006)}]{Kordyuk_ARPES}
\bibinfo{author}{\bibfnamefont{A.~A.} \bibnamefont{{Kordyuk}}}
  \bibnamefont{and} \bibinfo{author}{\bibfnamefont{S.~V.}
  \bibnamefont{{Borisenko}}}, \bibinfo{journal}{Low Temp. Phys.}
  \textbf{\bibinfo{volume}{32}}, \bibinfo{pages}{298} (\bibinfo{year}{2006}).

\bibitem[{\citenamefont{Hubbard}(1963)}]{Hubbard63}
\bibinfo{author}{\bibfnamefont{J.}~\bibnamefont{Hubbard}},
  \bibinfo{journal}{Proc. R. Soc. London, Ser. A}
  \textbf{\bibinfo{volume}{276}}, \bibinfo{pages}{238} (\bibinfo{year}{1963}).

\bibitem[{\citenamefont{Kanamori}(1963)}]{Kanamori63}
\bibinfo{author}{\bibfnamefont{J.}~\bibnamefont{Kanamori}},
  \bibinfo{journal}{Prog. Theor. Phys.} \textbf{\bibinfo{volume}{30}},
  \bibinfo{pages}{275} (\bibinfo{year}{1963}).

\bibitem[{\citenamefont{Gutzwiller}(1963)}]{Gutzwiller63}
\bibinfo{author}{\bibfnamefont{M.~C.} \bibnamefont{Gutzwiller}},
  \bibinfo{journal}{Phys. Rev. Lett.} \textbf{\bibinfo{volume}{10}},
  \bibinfo{pages}{159} (\bibinfo{year}{1963}).

\bibitem[{\citenamefont{{Husemann} and
  {Salmhofer}}(2009)}]{HusemannSalmhofer2009}
\bibinfo{author}{\bibfnamefont{C.}~\bibnamefont{{Husemann}}} \bibnamefont{and}
  \bibinfo{author}{\bibfnamefont{M.}~\bibnamefont{{Salmhofer}}},
  \bibinfo{journal}{Phys. Rev. B} \textbf{\bibinfo{volume}{79}},
  \bibinfo{pages}{195125} (\bibinfo{year}{2009}).

\bibitem[{\citenamefont{Husemann et~al.}(2012)\citenamefont{Husemann, Giering,
  and Salmhofer}}]{HGS2011}
\bibinfo{author}{\bibfnamefont{C.}~\bibnamefont{Husemann}},
  \bibinfo{author}{\bibfnamefont{K.-U.} \bibnamefont{Giering}},
  \bibnamefont{and}
  \bibinfo{author}{\bibfnamefont{M.}~\bibnamefont{Salmhofer}},
  \bibinfo{journal}{Phys. Rev. B} \textbf{\bibinfo{volume}{85}},
  \bibinfo{pages}{075121} (\bibinfo{year}{2012}).

\bibitem[{\citenamefont{Honerkamp and
  Salmhofer}(2001)}]{HonerkampSalmhofer_Tflowletter}
\bibinfo{author}{\bibfnamefont{C.}~\bibnamefont{Honerkamp}} \bibnamefont{and}
  \bibinfo{author}{\bibfnamefont{M.}~\bibnamefont{Salmhofer}},
  \bibinfo{journal}{Phys. Rev. Lett.} \textbf{\bibinfo{volume}{87}},
  \bibinfo{pages}{187004} (\bibinfo{year}{2001}).

\bibitem[{\citenamefont{{Honerkamp} and
  {Salmhofer}}(2001)}]{HonerkampSalmhofer_Tflow}
\bibinfo{author}{\bibfnamefont{C.}~\bibnamefont{{Honerkamp}}} \bibnamefont{and}
  \bibinfo{author}{\bibfnamefont{M.}~\bibnamefont{{Salmhofer}}},
  \bibinfo{journal}{Phys. Rev. B} \textbf{\bibinfo{volume}{64}},
  \bibinfo{pages}{184516} (\bibinfo{year}{2001}).

\bibitem[{\citenamefont{Hankevych et~al.}(2003)\citenamefont{Hankevych, Kyung,
  and Tremblay}}]{Hankevych2003}
\bibinfo{author}{\bibfnamefont{V.}~\bibnamefont{Hankevych}},
  \bibinfo{author}{\bibfnamefont{B.}~\bibnamefont{Kyung}}, \bibnamefont{and}
  \bibinfo{author}{\bibfnamefont{A.-M.~S.} \bibnamefont{Tremblay}},
  \bibinfo{journal}{Phys. Rev. B} \textbf{\bibinfo{volume}{68}},
  \bibinfo{pages}{214405} (\bibinfo{year}{2003}).

\bibitem[{\citenamefont{Rech et~al.}(2006)\citenamefont{Rech, P\'epin, and
  Chubukov}}]{Rech2006}
\bibinfo{author}{\bibfnamefont{J.}~\bibnamefont{Rech}},
  \bibinfo{author}{\bibfnamefont{C.}~\bibnamefont{P\'epin}}, \bibnamefont{and}
  \bibinfo{author}{\bibfnamefont{A.~V.} \bibnamefont{Chubukov}},
  \bibinfo{journal}{Phys. Rev. B} \textbf{\bibinfo{volume}{74}},
  \bibinfo{pages}{195126} (\bibinfo{year}{2006}).

\bibitem[{\citenamefont{Salmhofer}(1998)}]{Salmhofer_ContRenFerm}
\bibinfo{author}{\bibfnamefont{M.}~\bibnamefont{Salmhofer}},
  \bibinfo{journal}{Comm. Math. Phys.} \textbf{\bibinfo{volume}{194}},
  \bibinfo{pages}{249} (\bibinfo{year}{1998}).

\bibitem[{\citenamefont{Zanchi and Schulz}(1998)}]{ZanchiSchulz98}
\bibinfo{author}{\bibfnamefont{D.}~\bibnamefont{Zanchi}} \bibnamefont{and}
  \bibinfo{author}{\bibfnamefont{H.~J.} \bibnamefont{Schulz}},
  \bibinfo{journal}{Europhys. Lett.} \textbf{\bibinfo{volume}{44}},
  \bibinfo{pages}{235} (\bibinfo{year}{1998}).

\bibitem[{\citenamefont{Halboth and Metzner}(2000)}]{HalbothMetzner}
\bibinfo{author}{\bibfnamefont{C.~J.} \bibnamefont{Halboth}} \bibnamefont{and}
  \bibinfo{author}{\bibfnamefont{W.}~\bibnamefont{Metzner}},
  \bibinfo{journal}{Phys. Rev. B} \textbf{\bibinfo{volume}{61}},
  \bibinfo{pages}{7364} (\bibinfo{year}{2000}).

\bibitem[{\citenamefont{Honerkamp et~al.}(2001)\citenamefont{Honerkamp,
  Salmhofer, Furukawa, and Rice}}]{HonerkampSalmhofer_Umklapp}
\bibinfo{author}{\bibfnamefont{C.}~\bibnamefont{Honerkamp}},
  \bibinfo{author}{\bibfnamefont{M.}~\bibnamefont{Salmhofer}},
  \bibinfo{author}{\bibfnamefont{N.}~\bibnamefont{Furukawa}}, \bibnamefont{and}
  \bibinfo{author}{\bibfnamefont{T.~M.} \bibnamefont{Rice}},
  \bibinfo{journal}{Phys. Rev. B} \textbf{\bibinfo{volume}{63}},
  \bibinfo{pages}{035109} (\bibinfo{year}{2001}).

\bibitem[{\citenamefont{Karrasch et~al.}(2008)\citenamefont{Karrasch, Hedden,
  Peters, Pruschke, Sch\"onhammer, and Meden}}]{KarraschHeddenMeden2008}
\bibinfo{author}{\bibfnamefont{C.}~\bibnamefont{Karrasch}},
  \bibinfo{author}{\bibfnamefont{R.}~\bibnamefont{Hedden}},
  \bibinfo{author}{\bibfnamefont{R.}~\bibnamefont{Peters}},
  \bibinfo{author}{\bibfnamefont{T.}~\bibnamefont{Pruschke}},
  \bibinfo{author}{\bibfnamefont{K.}~\bibnamefont{Sch\"onhammer}},
  \bibnamefont{and} \bibinfo{author}{\bibfnamefont{V.}~\bibnamefont{Meden}},
  \bibinfo{journal}{J. Phys. Cond. Mat.} \textbf{\bibinfo{volume}{20}},
  \bibinfo{pages}{345205} (\bibinfo{year}{2008}).

\bibitem[{\citenamefont{Jakobs et~al.}(2010)\citenamefont{Jakobs, Pletyukhov,
  and Schoeller}}]{SeverinPletyukovSchoeller2010}
\bibinfo{author}{\bibfnamefont{S.~G.} \bibnamefont{Jakobs}},
  \bibinfo{author}{\bibfnamefont{M.}~\bibnamefont{Pletyukhov}},
  \bibnamefont{and}
  \bibinfo{author}{\bibfnamefont{H.}~\bibnamefont{Schoeller}},
  \bibinfo{journal}{Phys. Rev. B} \textbf{\bibinfo{volume}{81}},
  \bibinfo{pages}{195109} (\bibinfo{year}{2010}).

\bibitem[{\citenamefont{Karrasch et~al.}(2010)\citenamefont{Karrasch,
  Pletyukhov, Borda, and Meden}}]{KarraschEtal2010}
\bibinfo{author}{\bibfnamefont{C.}~\bibnamefont{Karrasch}},
  \bibinfo{author}{\bibfnamefont{M.}~\bibnamefont{Pletyukhov}},
  \bibinfo{author}{\bibfnamefont{L.}~\bibnamefont{Borda}}, \bibnamefont{and}
  \bibinfo{author}{\bibfnamefont{V.}~\bibnamefont{Meden}},
  \bibinfo{journal}{Phys. Rev. B} \textbf{\bibinfo{volume}{81}},
  \bibinfo{pages}{125122} (\bibinfo{year}{2010}).

\bibitem[{\citenamefont{Metzner et~al.}(2012)\citenamefont{Metzner, Salmhofer,
  Honerkamp, Meden, and Sch\"onhammer}}]{FRG_Overview}
\bibinfo{author}{\bibfnamefont{W.}~\bibnamefont{Metzner}},
  \bibinfo{author}{\bibfnamefont{M.}~\bibnamefont{Salmhofer}},
  \bibinfo{author}{\bibfnamefont{C.}~\bibnamefont{Honerkamp}},
  \bibinfo{author}{\bibfnamefont{V.}~\bibnamefont{Meden}}, \bibnamefont{and}
  \bibinfo{author}{\bibfnamefont{K.}~\bibnamefont{Sch\"onhammer}},
  \bibinfo{journal}{Rev. Mod. Phys.} \textbf{\bibinfo{volume}{84}},
  \bibinfo{pages}{299} (\bibinfo{year}{2012}).

\bibitem[{\citenamefont{Salmhofer and
  Honerkamp}(2001)}]{SalmhoferHonerkamp_ProgTheor}
\bibinfo{author}{\bibfnamefont{M.}~\bibnamefont{Salmhofer}} \bibnamefont{and}
  \bibinfo{author}{\bibfnamefont{C.}~\bibnamefont{Honerkamp}},
  \bibinfo{journal}{Prog. Theor. Phys.} \textbf{\bibinfo{volume}{105}},
  \bibinfo{pages}{1} (\bibinfo{year}{2001}).

\bibitem[{\citenamefont{Salmhofer et~al.}(2004)\citenamefont{Salmhofer,
  Honerkamp, Metzner, and Lauscher}}]{RGbrokenSym}
\bibinfo{author}{\bibfnamefont{M.}~\bibnamefont{Salmhofer}},
  \bibinfo{author}{\bibfnamefont{C.}~\bibnamefont{Honerkamp}},
  \bibinfo{author}{\bibfnamefont{W.}~\bibnamefont{Metzner}}, \bibnamefont{and}
  \bibinfo{author}{\bibfnamefont{O.}~\bibnamefont{Lauscher}},
  \bibinfo{journal}{Prog. Theor. Phys.} \textbf{\bibinfo{volume}{112}},
  \bibinfo{pages}{943} (\bibinfo{year}{2004}).

\bibitem[{\citenamefont{Gersch et~al.}(2005)\citenamefont{Gersch, Honerkamp,
  Rohe, and Metzner}}]{GerschEtal2005}
\bibinfo{author}{\bibfnamefont{R.}~\bibnamefont{Gersch}},
  \bibinfo{author}{\bibfnamefont{C.}~\bibnamefont{Honerkamp}},
  \bibinfo{author}{\bibfnamefont{D.}~\bibnamefont{Rohe}}, \bibnamefont{and}
  \bibinfo{author}{\bibfnamefont{W.}~\bibnamefont{Metzner}},
  \bibinfo{journal}{Eur. Phys. J. B} \textbf{\bibinfo{volume}{48}},
  \bibinfo{pages}{349} (\bibinfo{year}{2005}).

\bibitem[{\citenamefont{Strack et~al.}(2008)\citenamefont{Strack, Gersch, and
  Metzner}}]{StrackGerschMetzner08}
\bibinfo{author}{\bibfnamefont{P.}~\bibnamefont{Strack}},
  \bibinfo{author}{\bibfnamefont{R.}~\bibnamefont{Gersch}}, \bibnamefont{and}
  \bibinfo{author}{\bibfnamefont{W.}~\bibnamefont{Metzner}},
  \bibinfo{journal}{Phys. Rev. B} \textbf{\bibinfo{volume}{78}},
  \bibinfo{pages}{014522} (\bibinfo{year}{2008}).

\bibitem[{\citenamefont{Baier et~al.}(2004)\citenamefont{Baier, Bick, and
  Wetterich}}]{BaierBickWetterich04}
\bibinfo{author}{\bibfnamefont{T.}~\bibnamefont{Baier}},
  \bibinfo{author}{\bibfnamefont{E.}~\bibnamefont{Bick}}, \bibnamefont{and}
  \bibinfo{author}{\bibfnamefont{C.}~\bibnamefont{Wetterich}},
  \bibinfo{journal}{Phys. Rev. B} \textbf{\bibinfo{volume}{70}},
  \bibinfo{pages}{125111} (\bibinfo{year}{2004}).

\bibitem[{\citenamefont{Friederich et~al.}(2011)\citenamefont{Friederich,
  Krahl, and Wetterich}}]{Friederich2011}
\bibinfo{author}{\bibfnamefont{S.}~\bibnamefont{Friederich}},
  \bibinfo{author}{\bibfnamefont{H.~C.} \bibnamefont{Krahl}}, \bibnamefont{and}
  \bibinfo{author}{\bibfnamefont{C.}~\bibnamefont{Wetterich}},
  \bibinfo{journal}{Phys. Rev. B} \textbf{\bibinfo{volume}{83}},
  \bibinfo{pages}{155125} (\bibinfo{year}{2011}).

\bibitem[{\citenamefont{Feldman et~al.}(2000)\citenamefont{Feldman, Salmhofer,
  and Trubowitz}}]{FeldmanEtal2000}
\bibinfo{author}{\bibfnamefont{J.}~\bibnamefont{Feldman}},
  \bibinfo{author}{\bibfnamefont{M.}~\bibnamefont{Salmhofer}},
  \bibnamefont{and}
  \bibinfo{author}{\bibfnamefont{E.}~\bibnamefont{Trubowitz}},
  \bibinfo{journal}{Comm. Pure and Appl. Math.} \textbf{\bibinfo{volume}{53}},
  \bibinfo{pages}{1350} (\bibinfo{year}{2000}).

\bibitem[{\citenamefont{Salmhofer}(2007)}]{DynAdjProp}
\bibinfo{author}{\bibfnamefont{M.}~\bibnamefont{Salmhofer}},
  \bibinfo{journal}{Ann. Phys.} \textbf{\bibinfo{volume}{16}},
  \bibinfo{pages}{171} (\bibinfo{year}{2007}).

\bibitem[{\citenamefont{Igoshev et~al.}(2011)\citenamefont{Igoshev, Irkhin, and
  Katanin}}]{Katanin_SigmaFM}
\bibinfo{author}{\bibfnamefont{P.~A.} \bibnamefont{Igoshev}},
  \bibinfo{author}{\bibfnamefont{V.~Y.} \bibnamefont{Irkhin}},
  \bibnamefont{and} \bibinfo{author}{\bibfnamefont{A.~A.}
  \bibnamefont{Katanin}}, \bibinfo{journal}{Phys. Rev. B}
  \textbf{\bibinfo{volume}{83}}, \bibinfo{pages}{245118}
  (\bibinfo{year}{2011}).

\bibitem[{\citenamefont{Honerkamp}(2001)}]{Honerkamp_ScatRate}
\bibinfo{author}{\bibfnamefont{C.}~\bibnamefont{Honerkamp}},
  \bibinfo{journal}{Eur. Phys. J. B} \textbf{\bibinfo{volume}{21}},
  \bibinfo{pages}{81} (\bibinfo{year}{2001}).

\bibitem[{\citenamefont{Zanchi}(2001)}]{Zanchi_QuasiparticleWeight}
\bibinfo{author}{\bibfnamefont{D.}~\bibnamefont{Zanchi}},
  \bibinfo{journal}{Europhys. Lett.} \textbf{\bibinfo{volume}{55}},
  \bibinfo{pages}{376} (\bibinfo{year}{2001}).

\bibitem[{\citenamefont{Honerkamp and
  Salmhofer}(2003)}]{HonerkampSalmhofer_QuasiparticleWeight}
\bibinfo{author}{\bibfnamefont{C.}~\bibnamefont{Honerkamp}} \bibnamefont{and}
  \bibinfo{author}{\bibfnamefont{M.}~\bibnamefont{Salmhofer}},
  \bibinfo{journal}{Phys. Rev. B} \textbf{\bibinfo{volume}{67}},
  \bibinfo{pages}{174504} (\bibinfo{year}{2003}).

\bibitem[{\citenamefont{Katanin}(2009)}]{Katanin-TwoLoop}
\bibinfo{author}{\bibfnamefont{A.~A.} \bibnamefont{Katanin}},
  \bibinfo{journal}{Phys. Rev. B} \textbf{\bibinfo{volume}{79}},
  \bibinfo{pages}{235119} (\bibinfo{year}{2009}).

\bibitem[{\citenamefont{Rohe and Metzner}(2005)}]{RoheMetzner}
\bibinfo{author}{\bibfnamefont{D.}~\bibnamefont{Rohe}} \bibnamefont{and}
  \bibinfo{author}{\bibfnamefont{W.}~\bibnamefont{Metzner}},
  \bibinfo{journal}{Phys. Rev. B} \textbf{\bibinfo{volume}{71}},
  \bibinfo{pages}{115116} (\bibinfo{year}{2005}).

\bibitem[{\citenamefont{Katanin and Kampf}(2004)}]{KataninKampf2004}
\bibinfo{author}{\bibfnamefont{A.~A.} \bibnamefont{Katanin}} \bibnamefont{and}
  \bibinfo{author}{\bibfnamefont{A.~P.} \bibnamefont{Kampf}},
  \bibinfo{journal}{Phys. Rev. Lett.} \textbf{\bibinfo{volume}{93}},
  \bibinfo{pages}{106406} (\bibinfo{year}{2004}).

\bibitem[{\citenamefont{Katanin}(2004)}]{Katanin_WardId}
\bibinfo{author}{\bibfnamefont{A.~A.} \bibnamefont{Katanin}},
  \bibinfo{journal}{Phys. Rev. B} \textbf{\bibinfo{volume}{70}},
  \bibinfo{pages}{115109} (\bibinfo{year}{2004}).

\bibitem[{\citenamefont{Salmhofer}(1999)}]{Salmhofer_Book}
\bibinfo{author}{\bibfnamefont{M.}~\bibnamefont{Salmhofer}},
  \emph{\bibinfo{title}{Renormalization}} (\bibinfo{publisher}{Springer},
  \bibinfo{address}{Berlin}, \bibinfo{year}{1999}).

\bibitem[{Not()}]{Note1}
\bibinfo{note}{The present text and Ref.~\protect \citenum {HGS2011} differ by
  a conventional factor of two in the interaction vertex; the stopping
  condition is the same in both studies.}

\bibitem[{\citenamefont{Honerkamp et~al.}(2004)\citenamefont{Honerkamp, Rohe,
  Andergassen, and Enss}}]{HonerkampInteractionFlow2004}
\bibinfo{author}{\bibfnamefont{C.}~\bibnamefont{Honerkamp}},
  \bibinfo{author}{\bibfnamefont{D.}~\bibnamefont{Rohe}},
  \bibinfo{author}{\bibfnamefont{S.}~\bibnamefont{Andergassen}},
  \bibnamefont{and} \bibinfo{author}{\bibfnamefont{T.}~\bibnamefont{Enss}},
  \bibinfo{journal}{Phys. Rev. B} \textbf{\bibinfo{volume}{70}},
  \bibinfo{pages}{235115} (\bibinfo{year}{2004}).

\bibitem[{\citenamefont{Ortloff et~al.}()\citenamefont{Ortloff, Husemann,
  Honerkamp, and Salmhofer}}]{OrtloffHusemann}
\bibinfo{author}{\bibfnamefont{J.}~\bibnamefont{Ortloff}},
  \bibinfo{author}{\bibfnamefont{C.}~\bibnamefont{Husemann}},
  \bibinfo{author}{\bibfnamefont{C.}~\bibnamefont{Honerkamp}},
  \bibnamefont{and}
  \bibinfo{author}{\bibfnamefont{M.}~\bibnamefont{Salmhofer}},
  \bibinfo{note}{in preparation}.

\bibitem[{\citenamefont{Metzner et~al.}(2003)\citenamefont{Metzner, Rohe, and
  Andergassen}}]{MetznerRoheAndergassen2003}
\bibinfo{author}{\bibfnamefont{W.}~\bibnamefont{Metzner}},
  \bibinfo{author}{\bibfnamefont{D.}~\bibnamefont{Rohe}}, \bibnamefont{and}
  \bibinfo{author}{\bibfnamefont{S.}~\bibnamefont{Andergassen}},
  \bibinfo{journal}{Phys. Rev. Lett.} \textbf{\bibinfo{volume}{91}},
  \bibinfo{pages}{066402} (\bibinfo{year}{2003}).

\bibitem[{\citenamefont{Neumayr and Metzner}(2003)}]{Metzner2003_Nematicity}
\bibinfo{author}{\bibfnamefont{A.}~\bibnamefont{Neumayr}} \bibnamefont{and}
  \bibinfo{author}{\bibfnamefont{W.}~\bibnamefont{Metzner}},
  \bibinfo{journal}{Phys. Rev. B} \textbf{\bibinfo{volume}{67}},
  \bibinfo{pages}{035112} (\bibinfo{year}{2003}).

\bibitem[{\citenamefont{Yamase and Metzner}(2007)}]{Metzner2006_Nematicity}
\bibinfo{author}{\bibfnamefont{H.}~\bibnamefont{Yamase}} \bibnamefont{and}
  \bibinfo{author}{\bibfnamefont{W.}~\bibnamefont{Metzner}},
  \bibinfo{journal}{Phys. Rev. B} \textbf{\bibinfo{volume}{75}},
  \bibinfo{pages}{155117} (\bibinfo{year}{2007}).

\bibitem[{\citenamefont{Husemann and Metzner}(2012)}]{HusemannMetzner2012}
\bibinfo{author}{\bibfnamefont{C.}~\bibnamefont{Husemann}} \bibnamefont{and}
  \bibinfo{author}{\bibfnamefont{W.}~\bibnamefont{Metzner}},
  \bibinfo{journal}{Phys. Rev. B} \textbf{\bibinfo{volume}{86}},
  \bibinfo{pages}{085113} (\bibinfo{year}{2012}).

\bibitem[{\citenamefont{Halboth and Metzner}(1997)}]{HalbothMetzner97}
\bibinfo{author}{\bibfnamefont{C.~J.} \bibnamefont{Halboth}} \bibnamefont{and}
  \bibinfo{author}{\bibfnamefont{W.}~\bibnamefont{Metzner}},
  \bibinfo{journal}{Z. Phys. B} \textbf{\bibinfo{volume}{102}},
  \bibinfo{pages}{501} (\bibinfo{year}{1997}).

\bibitem[{\citenamefont{{Feldman} and
  {Salmhofer}}(2008)}]{FeldmanSalmhofer2008b}
\bibinfo{author}{\bibfnamefont{J.}~\bibnamefont{{Feldman}}} \bibnamefont{and}
  \bibinfo{author}{\bibfnamefont{M.}~\bibnamefont{{Salmhofer}}},
  \bibinfo{journal}{Rev. Math. Phys.} \textbf{\bibinfo{volume}{20}},
  \bibinfo{pages}{275} (\bibinfo{year}{2008}).

\bibitem[{\citenamefont{Schmidt and Enss}(2011)}]{SchmidtEnss}
\bibinfo{author}{\bibfnamefont{R.}~\bibnamefont{Schmidt}} \bibnamefont{and}
  \bibinfo{author}{\bibfnamefont{T.}~\bibnamefont{Enss}},
  \bibinfo{journal}{Phys. Rev. A} \textbf{\bibinfo{volume}{83}},
  \bibinfo{pages}{063620} (\bibinfo{year}{2011}).

\bibitem[{\citenamefont{Dzyaloshinskii}(1996)}]{Dzyaloshinskii1996}
\bibinfo{author}{\bibfnamefont{I.}~\bibnamefont{Dzyaloshinskii}},
  \bibinfo{journal}{J. Phys. I France} \textbf{\bibinfo{volume}{6}},
  \bibinfo{pages}{119} (\bibinfo{year}{1996}).

\bibitem[{\citenamefont{Metlitski and Sachdev}(2010)}]{MelitskiSachdev2010}
\bibinfo{author}{\bibfnamefont{M.~A.} \bibnamefont{Metlitski}}
  \bibnamefont{and} \bibinfo{author}{\bibfnamefont{S.}~\bibnamefont{Sachdev}},
  \bibinfo{journal}{Phys. Rev. B} \textbf{\bibinfo{volume}{82}},
  \bibinfo{pages}{075127} (\bibinfo{year}{2010}).

\bibitem[{\citenamefont{Mross et~al.}(2010)\citenamefont{Mross, McGreevy, Liu,
  and Senthil}}]{MrossEtal2010}
\bibinfo{author}{\bibfnamefont{D.~F.} \bibnamefont{Mross}},
  \bibinfo{author}{\bibfnamefont{J.}~\bibnamefont{McGreevy}},
  \bibinfo{author}{\bibfnamefont{H.}~\bibnamefont{Liu}}, \bibnamefont{and}
  \bibinfo{author}{\bibfnamefont{T.}~\bibnamefont{Senthil}},
  \bibinfo{journal}{Phys. Rev. B} \textbf{\bibinfo{volume}{82}},
  \bibinfo{pages}{045121} (\bibinfo{year}{2010}).

\bibitem[{\citenamefont{Drukier et~al.}(2012)\citenamefont{Drukier, Bartosch,
  Isidori, and Kopietz}}]{DrukierEtal2012}
\bibinfo{author}{\bibfnamefont{C.}~\bibnamefont{Drukier}},
  \bibinfo{author}{\bibfnamefont{L.}~\bibnamefont{Bartosch}},
  \bibinfo{author}{\bibfnamefont{A.}~\bibnamefont{Isidori}}, \bibnamefont{and}
  \bibinfo{author}{\bibfnamefont{P.}~\bibnamefont{Kopietz}},
  \bibinfo{journal}{Phys. Rev. B} \textbf{\bibinfo{volume}{85}},
  \bibinfo{pages}{245120} (\bibinfo{year}{2012}).

\end{thebibliography}

\end{document}